\documentclass{jfm}

\usepackage{graphicx}
\usepackage{natbib}

\ifCUPmtlplainloaded \else
  \checkfont{eurm10}
  \iffontfound
    \IfFileExists{upmath.sty}
      {\typeout{^^JFound AMS Euler Roman fonts on the system,
                   using the 'upmath' package.^^J}%
       \usepackage{upmath}}
      {\typeout{^^JFound AMS Euler Roman fonts on the system, but you
                   dont seem to have the}%
       \typeout{'upmath' package installed. JFM.cls can take advantage
                 of these fonts,^^Jif you use 'upmath' package.^^J}%
       \providecommand\upi{\pi}%
      }
  \else
    \providecommand\upi{\pi}%
  \fi
\fi

\ifCUPmtlplainloaded \else
  \checkfont{msam10}
  \iffontfound
    \IfFileExists{amssymb.sty}
      {\typeout{^^JFound AMS Symbol fonts on the system, using the
                'amssymb' package.^^J}%
       \usepackage{amssymb}%
         \let\leq=\leqslant
         \let\geq=\geqslant
      }{}
  \fi
\fi

\ifCUPmtlplainloaded \else
  \IfFileExists{amsbsy.sty}
    {\typeout{^^JFound the 'amsbsy' package on the system, using it.^^J}%
     \usepackage{amsbsy}}
    {\providecommand\boldsymbol[1]{\mbox{\boldmath $##1$}}}
\fi

\providecommand\bnabla{\boldsymbol{\nabla}}
\providecommand\bcdot{\boldsymbol{\cdot}}

\newsavebox{\astrutbox}
\sbox{\astrutbox}{\rule[-5pt]{0pt}{20pt}}

\newcommand\sthalf{\ensuremath{{\scriptstyle\frac{3}{2}}}}
\newcommand\sth{\ensuremath{^{\sthalf}}}

\newcommand\thalf{\ensuremath{{\textstyle\frac{1}{2}}}}

\newcommand\blam{\boldsymbol{\lambda}}

\newcommand\uu{\boldsymbol{u}}
\newcommand\LL{\boldsymbol{L}}

\newcommand\hr{\boldsymbol{r}}

\newcommand{\cC}{\ensuremath{\mathcal{C}}}

\newcommand{\cW}{\ensuremath{\mathcal{W}}}

\providecommand\der{{\rm d}}
\providecommand\Der{{\rm D}}
\providecommand\pa{{\partial}}

\def\dom{ \,\textrm{d} \Omega}

\def\f#1#2{\frac{#1}{#2}}

\def\pp#1#2{\frac{\pa  #1}{\pa  #2}}
\def\dd#1#2{\frac{\der #1}{\der #2}}
\def\DD#1#2{\frac{\Der #1}{\Der #2}}

\def\ltw{{\hbox{${\lower 3pt\hbox{$<$}}\atop{\raise 1pt\hbox{$\sim$}}$}\,}}
\def\gtw{{\hbox{${\lower 3pt\hbox{$>$}}\atop{\raise 1pt\hbox{$\sim$}}$}\,}}

\title[On the late-time behaviour of a bounded, inviscid two-dimensional flow]{On the late-time behaviour of a bounded, inviscid two-dimensional flow}

\author[D. G. Dritschel, W. Qi and J. B. Marston]%
{David G. Dritschel$^1$%
  \thanks{Email address for correspondence: david.dritschel@st-andrews.ac.uk},\ns
Wanming Qi$^{2,3,4}$ and J. B. Marston$^{3,4}$}

\affiliation{$^1$School of Mathematics and Statistics, University of St 
Andrews, St Andrews KY16 9SS, UK\\[\affilskip]
$^2$School of Physics, Korea Institute for Advanced Study, 
Seoul 130-722, Korea\\[\affilskip]
$^3$Kavli Institute for Theoretical Physics, University of California, 
Santa Barbara, CA 93106-4030 USA\\[\affilskip]
$^4$Department of Physics, Brown University, Providence, RI 
02912-1843, USA}

\pubyear{2015}
\volume{???}
\pagerange{???--???}
\date{?; revised ?; accepted ?. - To be entered by editorial office}
\begin{document}

\maketitle

\begin{abstract}
Using complementary numerical approaches at high resolution, we study
the late-time behaviour of an inviscid, incompressible two-dimensional
flow on the surface of a sphere.  Starting from a random initial
vorticity field comprised of a small set of intermediate wavenumber spherical 
harmonics, we find that --- contrary to the predictions of equilibrium
statistical mechanics --- the flow does not evolve into a large-scale
steady state.  Instead, significant unsteadiness persists,
characterised by a population of persistent small-scale vortices
interacting with a large-scale oscillating quadrupolar vorticity field.  Moreover,
the vorticity develops a stepped, staircase distribution, consisting
of nearly homogeneous regions separated by sharp gradients.  The
persistence of unsteadiness is explained by a simple point vortex
model characterising the interactions between the four main vortices
which emerge.
\end{abstract}


\section{Introduction}
Two-dimensional flow models have often been used as an idealised
description of atmospheric and oceanic fluid dynamics.  These models
afford the maximum simplicity while retaining the key advective
nonlinearity of fluid motion and the associated scale cascade.  
However, such models fail to capture the intrinsic three-dimensionality
of atmospheric and oceanic flows at `small' scales, where the effects
of rotation and stratification strongly compete \citep{dtja99}.
Nevertheless, mathematically, much interest remains in understanding
how nonlinearity organises fluid motion in this simple context.

In this paper, we examine the behaviour of a bounded two-dimensional
incompressible inviscid (dissipationless) flow at long times $t$.
The flow evolves freely from some prescribed initial
condition, which is subsequently transformed by nonlinearity.  No
forcing is applied.  

This problem has been the subject of much previous research.
Almost all laboratory experiments and numerical simulations have
indicated a final state consisting of a steady large-scale circulation
pattern filling the whole domain (see
\citet{matthaeus91a,matthaeus91b,montgomery92,montgomery93,brands97,bouchet12}
for a sample of the vast literature). 
Unsteady final states have also been reported in 
several numerical simulations 
for doubly-periodic 
planar domains.
Segre and Kida found various unsteady final states using initial configurations of constant-vorticity patches \citep{SK98}, but the simulations appear not to have gone far enough in time for the proper final state to evolve \citep{YMC03}. Morita found a late-time persistent oscillatory state by starting from a perturbed zonal flow \citep{morita11}.  
However both examples of unsteady final states start from special initial conditions that may not develop broad-band turbulence in which energy is widely shared among many wavenumbers. It is commonly believed that although various metastable states can persist for a long time due to insufficient mixing, the flow will settle into a steady final state if it undergoes strong turbulent mixing during the evolution (for example see \citet{YMC03}). 
Theoretically, an explanation
for this has been proposed by \citet{miller90,miller92} and
\citet{robert91a,robert91b}, hereafter `MRS'.  (Reviews may be found in
\citet{chavanis02,eyink06,majda06,chavanis09,bouchet12,herbert13,qi14b}.)  The MRS
theory predicts that the nonlinear dynamics, characterised by
intense turbulent mixing, ultimately results in a stationary
large-scale, domain-filling flow.  Moreover, this flow is obtained as
a statistical mechanical equilibrium (assuming random fluctuations of
fine-scale vorticity) dependent only on the inviscid invariants.
Those include energy, circulation, and in general an infinite set of
`Casimirs' that determine the occupation of vorticity $\omega$ of each
level within the fluid (this is the `vorticity measure').

Verification of the MRS theory has met with mixed success
\citep{qi14b}.  The theory itself is difficult to work with when many
Casimirs are taken into account, and various simplifications have been
put forth (see e.g.\ the review by \citet{bouchet12}).  
Verification is also problematic when nearly all of the numerical methods used
suffer from some form of dissipation, artificial or real, not present
in the original inviscid system for which the theory was developed. 
A few numerical methods have been developed that conserve many
invariants, including the Casimirs \citep{abramov03,dubinkina10}. 
Note that as numerical simulations inevitably have finite resolution, the Casimirs or the vorticity field observed numerically are only the {\it coarse-grained} correspondence of those quantities of the inviscid system: the resolved vorticity field corresponds to the coarse-grained vorticity $\bar{\omega}$, not $\omega$, and the second- and higher-order Casimirs of the resolved vorticity are different from the original fine-grained ones dependent on $\omega$. 
These conservative numerical methods
would appear to be ideal, and moreover they would overcome the
long-standing problem of how unresolved scales impact resolved ones,
the `closure problem'.  However, an irrefutable feature of
two-dimensional turbulence is the enstrophy cascade, occurring as thin
vorticity filaments stretch and thin to ever finer scales.  By
conservation, this cascade is necessary to build up energy at
ever larger scales through an `inverse cascade' 
\citep[\& refs.]{fjortoft53,dritschel09a}, 
as indeed is anticipated in the MRS theory.  
That is, energy growth at large scales requires a cascade of
enstrophy to small scales.  Since the Casimirs involve integrals of
powers of the vorticity distribution, all but the first power (the
circulation) cannot be even approximately conserved at any finite
resolution, after sufficiently long time.  Forcing conservation of
quantities which cannot be conserved can only lead to spurious
conclusions.  

Verification of the MRS theory is further problematic because the invariant {\it fine-grained} Casimirs that the theory depends on cannot be measured from any numerical simulation of finite resolution. However in practice the resolved Casimirs are often used to determine the values of the fine-grained Casimirs (at most approximately).  
Notably, the coarse-grained
vorticity field $\bar\omega$ predicted by the MRS theory often agrees with that emerging at large times in numerical simulations, if the late-time values of the resolved Casimirs are applied.  Still these resolved Casimirs are not conserved and cannot be predicted from an
arbitrary initial flow state of a numerical simulation, which limits the predicative power of the MRS theory.

Another fundamental issue concerns ergodicity, namely that ensemble
averages over equi-probable states with the same conserved quantities
equate to time averages of a single state over a sufficiently long
period of time.  Two-dimensional flows are generally not ergodic
\citep{whitaker94,bricmont01,bouchet10}, though they are assumed to be in the MRS
theory.  It has been argued that MRS theory may still apply to states
which exhibit strong turbulent mixing over the course of their
evolution, e.g.\ to initially small-scale isotropic vorticity fields.
There are many such states.  But it is unclear whether they satisfy
ergodicity even approximately.

The present paper is not intended to be a critique only of MRS theory;
rather, we wish to present fresh evidence that even the central claim
for the existence of a final steady equilibrium vorticity distribution
$\bar\omega$ is likely to be false. Instead of the much studied doubly periodic planar domain, we consider flows on the sphere where statistical mechanics has not seen much application.
While we agree that one can never
carry out numerical simulations to infinite time, our results strongly
suggest that, at least for flows on a sphere, 
unsteadiness persists for all time.
Note that both \citet{SK98} and \citet{morita11} have focused on the type of final unsteadiness that is a result of special initial conditions (or insufficiently developed turbulence), but we argue that unsteadiness is generic by using a random initial state that generates fairly strong turbulent mixing: the unsteadiness is intrinsic to a system of compact coherent vortices. 
Our simulations are carried out much further in time and at much higher resolutions compared to the two studies, and the robustness of our result is verified by comparing two very different numerical approaches with fundamentally different forms of numerical dissipation. 

The paper is organised as follows.  In the next section, we briefly
review the equations and the associated conserved quantities, 
then describe two very different numerical methods employed.  Both 
methods are used in \S3 at many different resolutions, on the same
initial data, to study the long-time convergence of the results.
A few conclusions are offered in \S4.

\section {Governing equations and numerical methods}
\label{sec:setup}

\subsection {Two-dimensional flow on a sphere}
\label{sec:2dflow}

We wish to 
study a freely-evolving incompressible two-dimensional flow, with
no forcing or dissipation, confined to the surface of a unit sphere
centered at the origin.  In this case, the scalar vorticity $\omega$
normal to the surface is materially conserved,
\begin{equation}
\DD{\omega}{t} = \pp{\omega}{t}+\uu\bcdot\bnabla\omega=0 \,,
\label{voreq1}
\end{equation}
where $\uu=\hr\times\bnabla\psi$ is the incompressible velocity field,
$\hr$ is the position of a point on the surface ($|\hr|=1$), and
$\psi$ is the scalar streamfunction.  With $\bnabla\bcdot\uu=0$, the
definition $\omega=\hr\bcdot(\bnabla\times\uu)$ yields Poisson's
equation for $\psi$:
\begin{equation}
\nabla^2\psi = \omega \,.
\label{vorinv}
\end{equation}
These equations were originally derived by \cite{zermelo02} and
\cite{charney49}.

\subsection {Constants of the motion}
\label{sec:conservation}

Inviscid two-dimensional flow is an infinite-order Hamiltonian system
with the kinetic energy
\begin{equation}
E = \thalf\int\!\!\int |\uu|^2\dom = -\thalf\int\!\!\int \omega\psi\dom
\label{kinetic}
\end{equation}
serving as the Hamiltonian.  Above, the second expression is obtained
after integration by parts, and $\dom$ is the differential surface
area element.  The integral is taken over the entire surface of the
sphere.  Besides $E$, rotational symmetry about any axis implies that
the vector angular impulse
\begin{equation}
\LL = \int\!\!\int \omega\hr\dom
\label{angimp}
\end{equation}
is also conserved.  Non-zero $\LL$ implies a mean rotation with vector
angular velocity $3\LL/8\upi$.  (Below, we only consider flows with
$\LL=0$ to avoid starting with a flow organised at the largest
possible scale.)  Finally, material conservation of vorticity and
incompressibility imply a generally infinite number of additional
constraints, namely 
\begin{equation}
\cC_n = \int\!\!\int \omega^n\dom\,, \qquad n=2,\,3,\,4,\,...
\label{casimirs}
\end{equation}
are conserved. ($\cC_1=0$ automatically; no net circulation is
possible on a closed surface).  The $\cC_n$ are called the `Casimirs'.
Their conservation is a consequence of the fact that the fractional
area occupied by each vorticity value is conserved.  For a continuous
vorticity distribution, this fractional area is generally
infinitesimal, and it is then useful to describe it as a `measure', or
probability density $p(\omega)$: the area occupied by vorticity in the
range $[\omega,\omega+\der\omega]$ is $4\pi p(\omega)\der\omega$.  The
conservation of all $\cC_n$ is equivalent to the conservation of
$p(\omega)$.

The quadratic Casimir $\cC_2$ is proportional to the `enstrophy' $Z$ 
(i.e.\ $Z=\cC_2/2$).  It
plays a key role in the so called `dual cascade' of two-dimensional
turbulence, where nonlinearity generally results in a direct cascade
of the spectral enstrophy density, enabling an inverse cascade of the
spectral energy density \citep{fjortoft53}.  This inverse cascade is often associated
with the emergence of large-scale flow features, a general prediction of the
MRS theory discussed in \S1.

Beyond the conservation of an infinite number of Casimirs,
Eq.~(\ref{voreq1}) further implies the existence of a topological constraint
that contours of vorticity cannot cross.  This nonlocal constraint is 
usually ignored in the construction of statistical descriptions. 

\subsection {Numerical methods}
\label{sec:numerical}

Two very different numerical methods are used to cross verify our
conclusions.  The first is a purely grid-based method using a
quasi-isotropic spherical `geodesic' grid consisting of $D$ cells --- hereafter
referred to as the `geodesic grid method' or `GGM' 
\citep{heikes95a,heikes95b,qi14b}.  In the absence of explicit numerical
dissipation, the GGM conserves both energy $E$ and enstrophy $Z$.
To remove enstrophy cascading to small scales, quad-harmonic
hyperviscosity is applied by adding the term
$\nu_4(\nabla^2+2)\nabla^6\omega$ to the right hand side of
(\ref{voreq1}) (the $(\nabla^2+2)$ operator ensures $\LL$ is
conserved).  The coefficient $\nu_4$ for any given resolution $D$ is
chosen so that the most rapidly dissipating mode decays at a rate of
$160$ (initially, $|\omega|_{\mathsf{max}}=5.8$, see below).  Adding
explicit dissipation of course changes the original problem, but we
argue it is necessary to account for the forward cascade of enstrophy
from resolved to unresolved scales.  Hyperviscosity, however, has
unwanted side effects, and in particular it can 
amplify vorticity extrema \citep{mariotti94,jimenez94,yao95}.
Nonetheless, hyperviscosity is more scale selective than ordinary
viscosity and much less dissipative overall for a given resolution
$D$.  Time stepping is carried out with a second-order leapfrog
scheme, using a time step of $\Delta t = 0.005$ at all resolutions.  
(The time step respects the CFL stability constraint.)  
Decorrelation of even and odd time levels in the
leapfrog scheme is avoided by applying a weak Robert-Asselin time
filter ($\alpha=0.001$), improved as recommended in \cite{williams09},
with the modification parameter $0.53$.  We have checked that using a smaller
value of $\alpha=10^{-4}$ makes little difference.  
Hyperviscosity is implemented explicitly.  
GGM is freely available online.\footnote{GGM 
is implemented in the application ``GCM'' that is available for 
OS X 10.9 and higher on the Apple Mac App Store at URL 
http://appstore.com/mac/gcm}

The second method is the `Combined Lagrangian Advection Method' (CLAM,
\cite{dritschel10}).  This is a hybrid method, using material
vorticity contours (solving $\der{\hr}/\der{t}=\uu(\hr,t)$) to
accurately and straightforwardly satisfy the conservative form of
(\ref{voreq1}), but also making use of conventional grid-based
procedures to invert (\ref{vorinv}) and to interpolate $\uu$ at
contour nodes.  CLAM is an extension of the closely related
Contour-Advective Semi-Lagrangian algorithm (CASL,
\cite{dritschel97}).  CLAM additionally solves (\ref{voreq1}) for
$\omega$ on a grid and blends this grid-based solution with the
contour-based one at the end of every time step to significantly
improve energy conservation compared to CASL.  This is essential for
carrying out exceptionally long simulations, as required here.  
In contrast to GGM which represents the vorticity as a continuous variable on a discrete grid in space, in CLAM the vorticity
itself is discretized, and the contours can move continuously in space.  
A contour spacing of $\Delta\omega=0.1$ is used in all simulations here,
corresponding to 109 contour levels.
Details of the method along with tests and comparisons with other
methods may be found in \cite{dritschel10}.

What is not described there is the specific implementation of CLAM in
spherical geometry.  Following \cite{moheb07}, we employ a regular
latitude-longitude grid of dimensions $n_g \times 2n_g$, enabling FFTs
in longitude.  Polar points are avoided by placing latitudes at
$-\upi/2 + (j-1/2)\upi/n_g$, for $j=1,\,2,\,...\,n_g$.  Latitude
derivatives are computed spectrally after carrying out FFTs along
great circles passing through the poles.  The inversion of Poisson's
equation (\ref{vorinv}) for $\psi$ is performed using 4th-order
compact differencing in latitude for each longitudinal wavenumber.
For the time evolution, CLAM evolves side-by-side a contour
representation of $\omega$, a gridded representation, and a residual
gridded field to minimise numerical dissipation and to achieve
excellent conservation properties (see \cite{dritschel10}).  The
gridded fields are evolved semi-spectrally, using the pseudo-spectral
method whereby nonlinear products are carried out in physical space.
Only the small residual field is damped by a weak bi-harmonic
hyperviscosity, with a damping rate of $2\omega_{\mathsf{rms}}(t)$ on
wavenumber $n_g$ (initially, $\omega_{\mathsf{rms}}=1.7060324$ and
decays thereafter).  A 4th-order Runge-Kutta method is used for time
stepping, with an adaptive time step $\Delta{t}=$
min$\{0.1\upi/|\omega|_{\mathsf{max}},0.7\upi/(n_g|\uu|_{\mathsf{max}})\}$.
Contours are regularised periodically by `contour surgery' at the
subgrid scale $\upi/16n_g$ \citep{dritschel88}, and occasionally by
contour regeneration, using the standardised parameter settings set
out in \cite{fontane09}.

Notably, the effective resolution of CLAM is approximately 
$16$ times finer in each direction than the basic 
`inversion' grid of dimensions $2n_g \times n_g$,
based on previous direct comparisons with standard pseudo-spectral 
methods \citep{dritschel09b,dritschel10,dritschel12}.

\begin{figure}
\begin{center}

$\begin{array}{c@{\hspace{0.12in}}c}
\multicolumn{1}{l}{\mbox{(a)}} &
\multicolumn{1}{l}{\mbox{(b)}}\\ [-0.0in]
\includegraphics[width = 2.4in]
{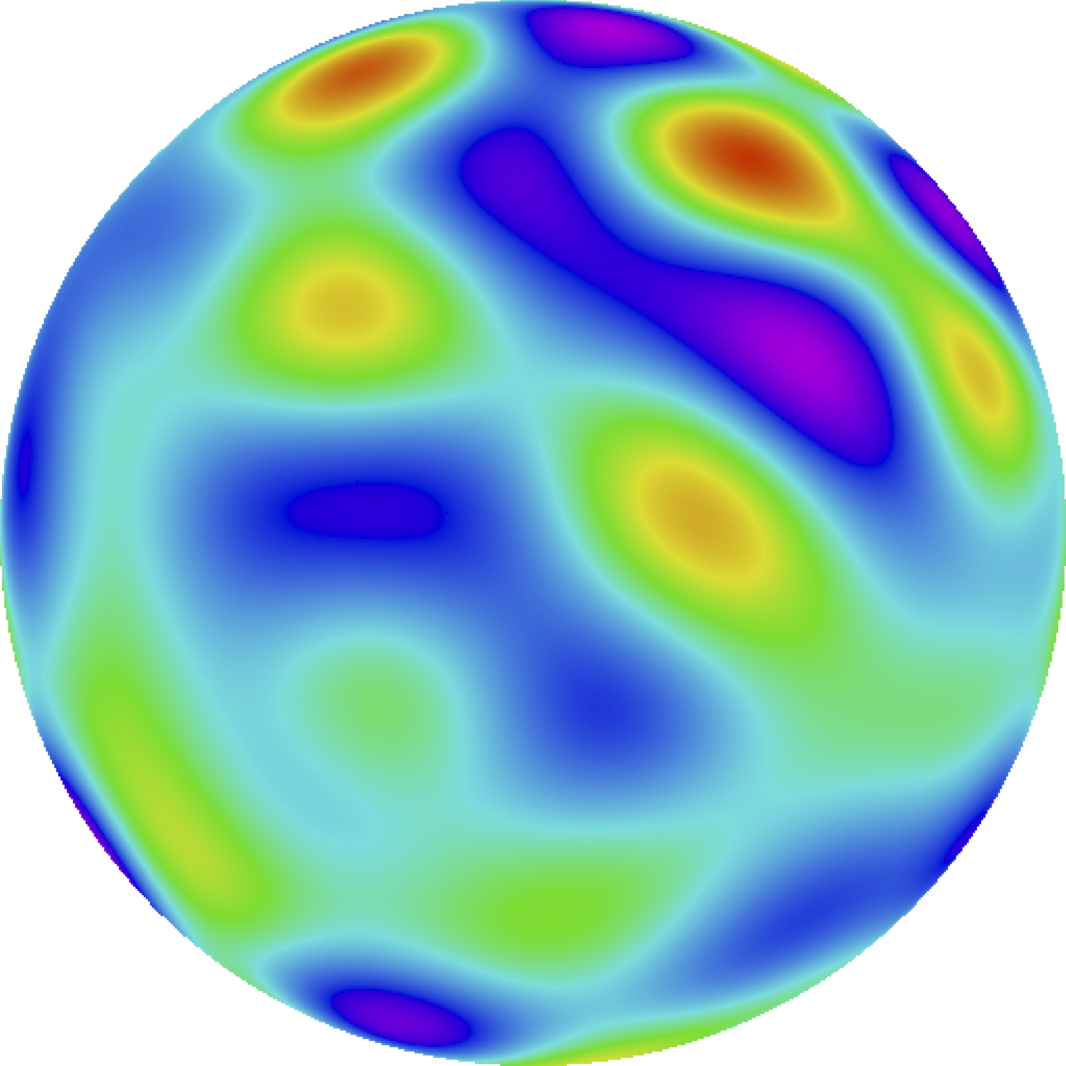} &
\includegraphics[width = 2.4in]
{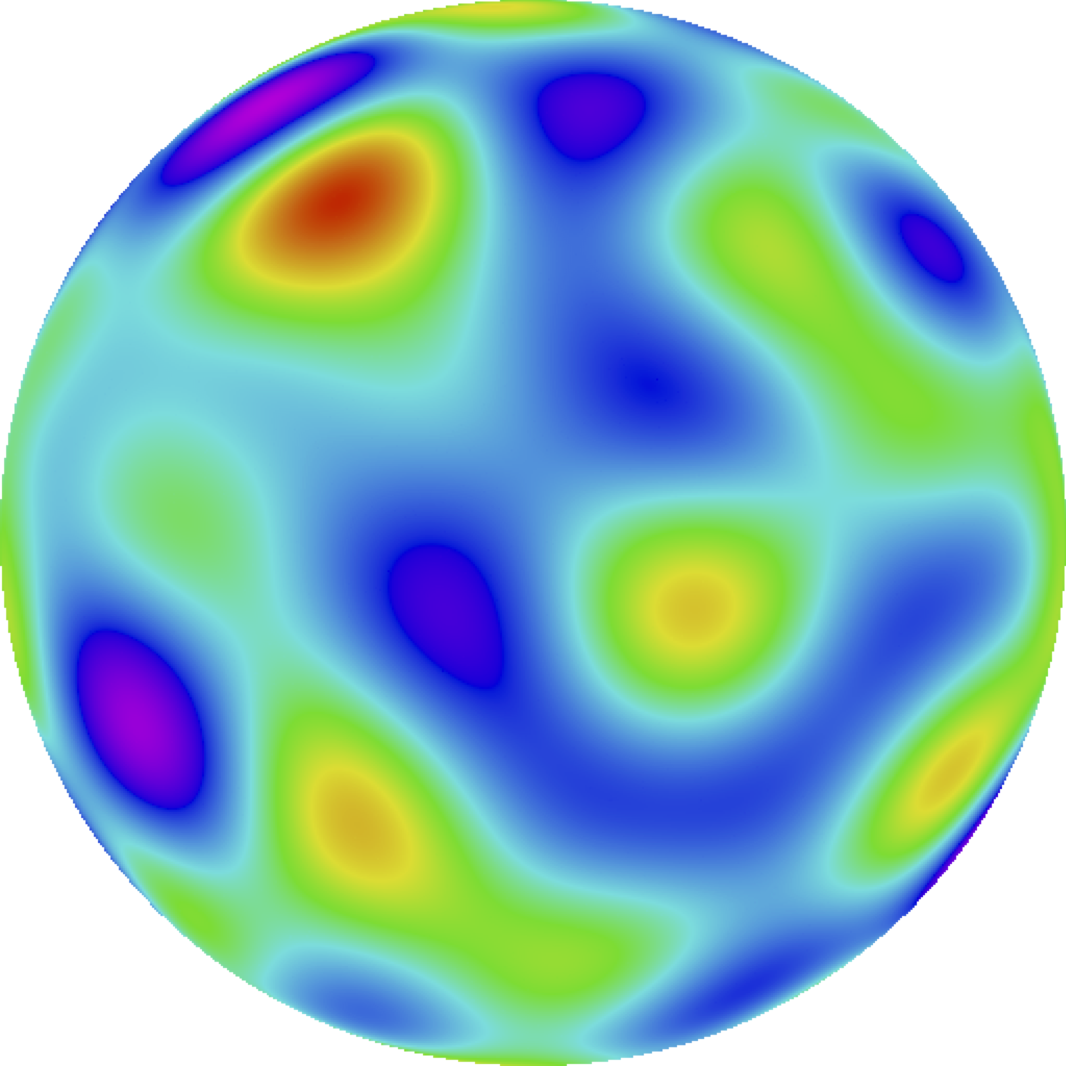}
\includegraphics[width = 0.35in]
{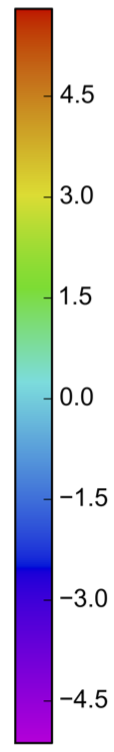}
\end{array}$

\end{center}
\caption{Initial vorticity distribution $\omega(\hr,0)$ as seen from
  the equatorial plane (a) at $0^{\circ}$ longitude, and (b) at
  $180^{\circ}$ longitude.  Colours (online) range from blue/purple for
  $\omega=\omega_{\mathsf{min}}=-5.1187$ through green to red for
  $\omega=\omega_{\mathsf{max}}=5.7996$.  An orthographic projection
  is used.}
\label{fig-vor0}
\end{figure}

\section {Results}
\label{sec:results}

\subsection {Flow initialisation and evolution}
\label{sec:init}

\begin{figure}
\begin{center}

$\begin{array}{c@{\hspace{0.12in}}c}
\multicolumn{1}{l}{\mbox{(a)}} &
\multicolumn{1}{l}{\mbox{(b)}}\\ [-0.0in]
\includegraphics[width = 2.56in]
{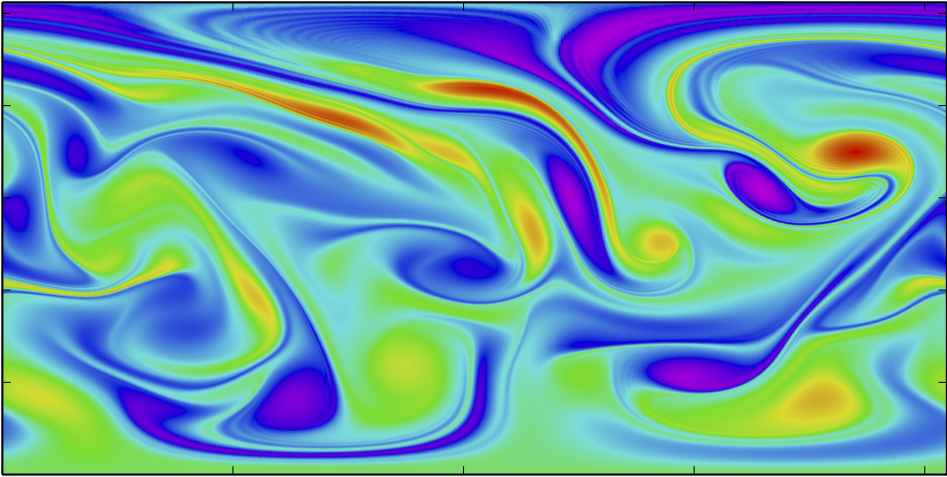} &
\includegraphics[width = 2.56in]
{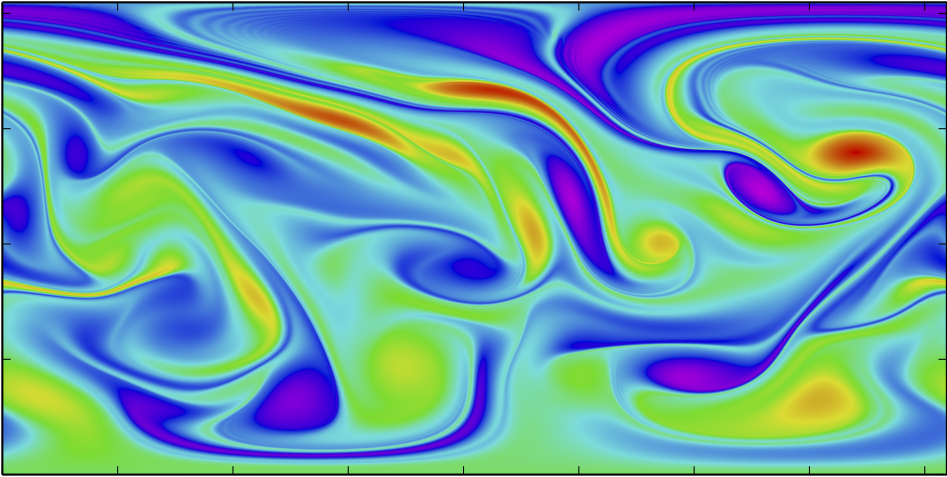}
\end{array}$

\vspace{0.1cm}

$\begin{array}{c@{\hspace{0.12in}}c}
\multicolumn{1}{l}{\mbox{(c)}} &
\multicolumn{1}{l}{\mbox{(d)}}\\ [-0.0in]
\includegraphics[width = 2.56in]
{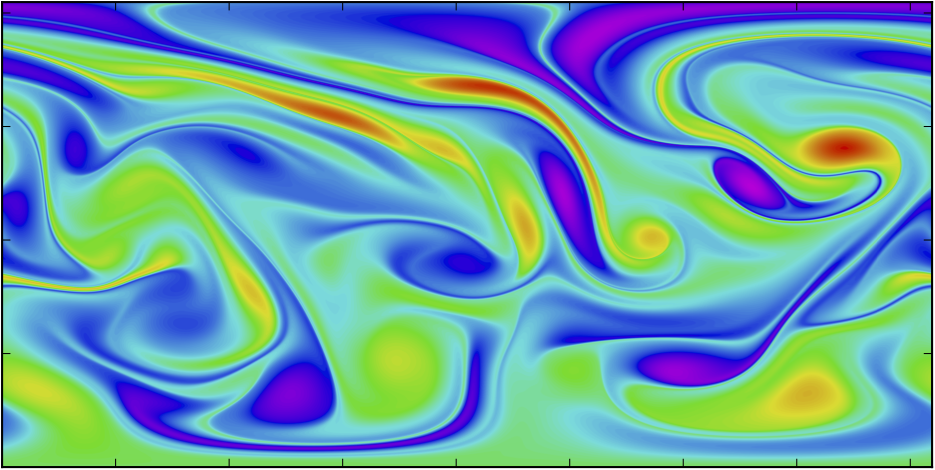} &
\includegraphics[width = 2.56in]
{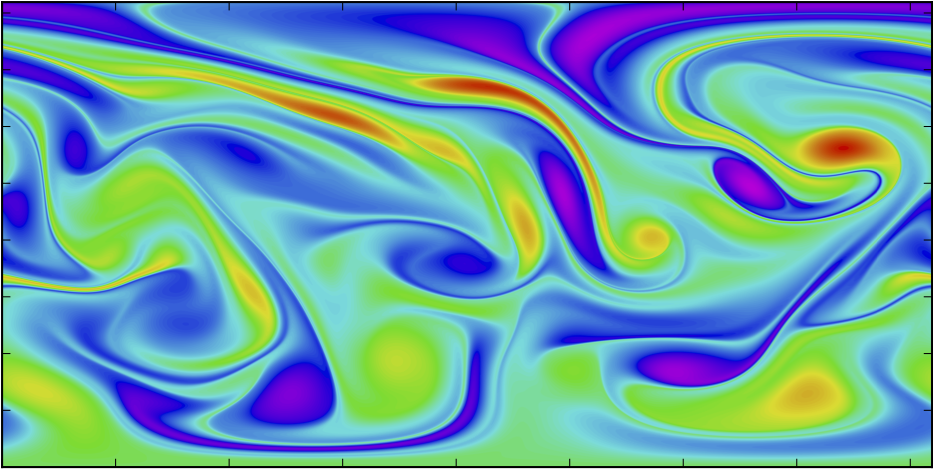}
\end{array}$

\end{center}

\caption{Early-time vorticity distribution $\omega(\hr,4)$ shown as a
  function of longitude and latitude for (a) GGM at $D=655362$, (b)
  GGM at $D=2621442$, (c) CLAM at $n_g=128$, and (d) CLAM at $n_g=256$.
  The images are rendered as described in figure~\ref{fig-vor0}.  The
  percentage decay in enstrophy $Z$ by this time is 0.13\%, 0.02\%,
  0.48\% and $<0.01\%$, respectively.  Same color scale as in Fig. \ref{fig-vor0}.}
\label{fig-vor4}
\end{figure}

A number of simulations differing only in resolution were carried out
using both the GGM and the CLAM methods.  In GGM, we used resolutions
$D=163842$, $655362$ and $2621442$, while in CLAM, we used resolutions
$n_g=128$, $256$, $512$ and $1024$.  All simulations start with an
identical vorticity field $\omega(\hr,0)$ built from a set of
spherical harmonics with degree $4\leq l \leq 10$; this field is shown
in figure~\ref{fig-vor0} (see appendix A for the amplitudes used).
Other initial conditions were tried and produce qualitatively similar
results.  The subsequent evolution, for early times, is nearly
indistinguishable in all of the numerical simulations conducted ---
this is illustrated in figure~\ref{fig-vor4} at $t=4$ (note, the eddy
turnover time
$t_{\mathsf{eddy}}=4\pi/|\omega|_{\mathsf{max}}=2.1668$).  Here, the
two highest resolution simulations carried out using GGM are compared
with the two lowest resolution simulations carried out using
CLAM.  By this stage of the evolution, the scale cascade of $\omega$
is still well resolved, though a small fraction of the enstrophy $Z$
($<1\%$) has already been lost.

\begin{figure}
\begin{center}

$\begin{array}{c@{\hspace{0.12in}}c}
\multicolumn{1}{l}{\mbox{(a)}} &
\multicolumn{1}{l}{\mbox{(b)}}\\ [-0.0in]
\includegraphics[width = 2.56in]
{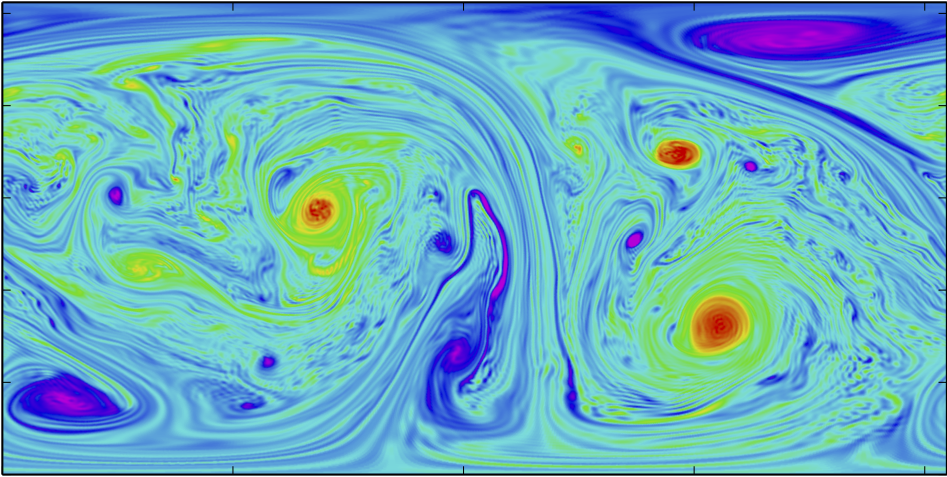} &
\includegraphics[width = 2.56in]
{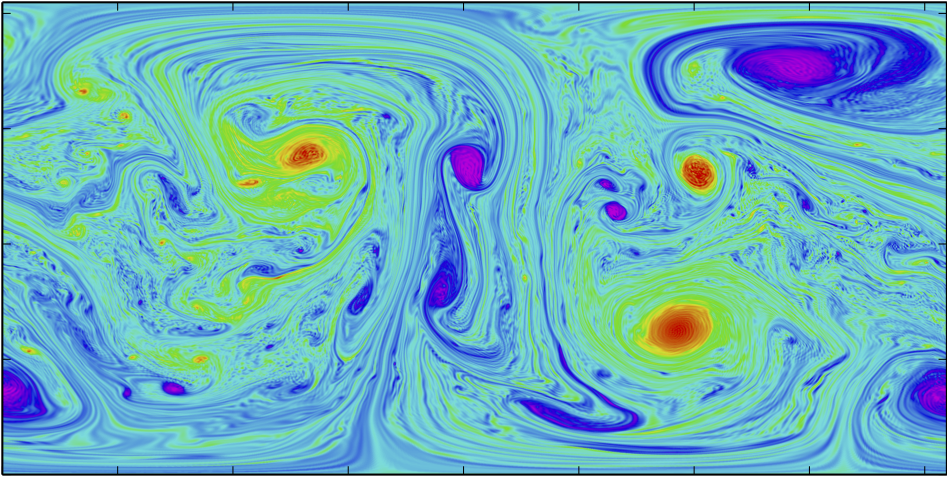}
\end{array}$

\vspace{0.1cm}

$\begin{array}{c@{\hspace{0.12in}}c}
\multicolumn{1}{l}{\mbox{(c)}} &
\multicolumn{1}{l}{\mbox{(d)}}\\ [-0.0in]
\includegraphics[width = 2.56in]
{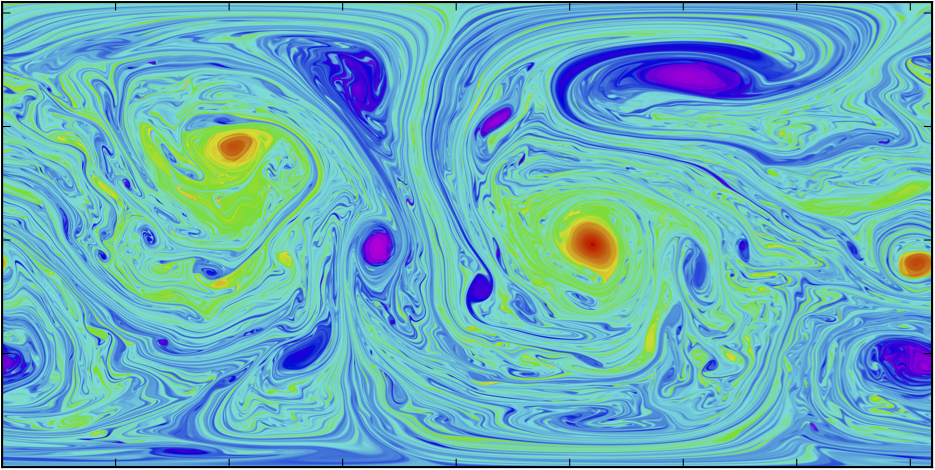} &
\includegraphics[width = 2.56in]
{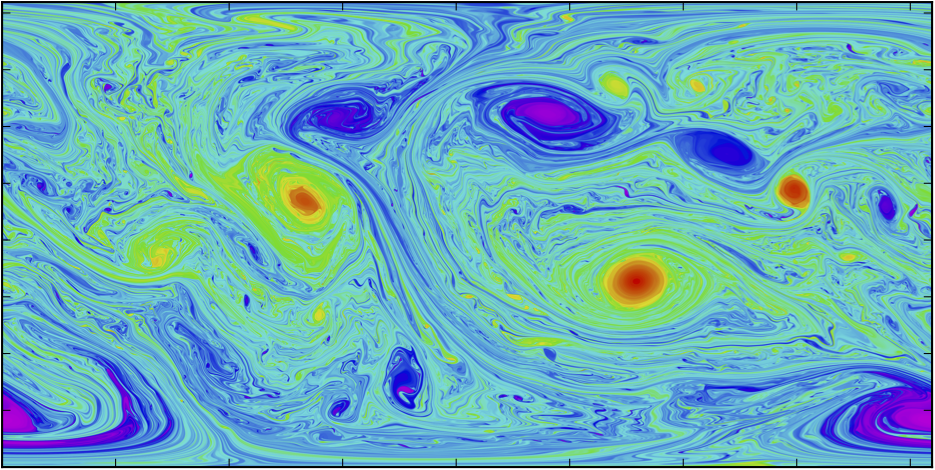}
\end{array}$

\vspace{0.1cm}

$\begin{array}{c@{\hspace{0.12in}}c}
\multicolumn{1}{l}{\mbox{(e)}} &
\multicolumn{1}{l}{\mbox{(f)}}\\ [-0.0in]
\includegraphics[width = 2.56in]
{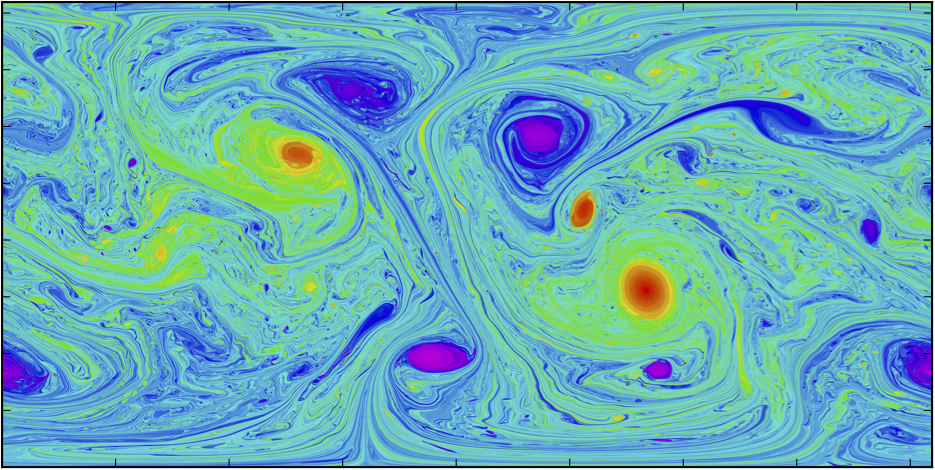} &
\includegraphics[width = 2.56in]
{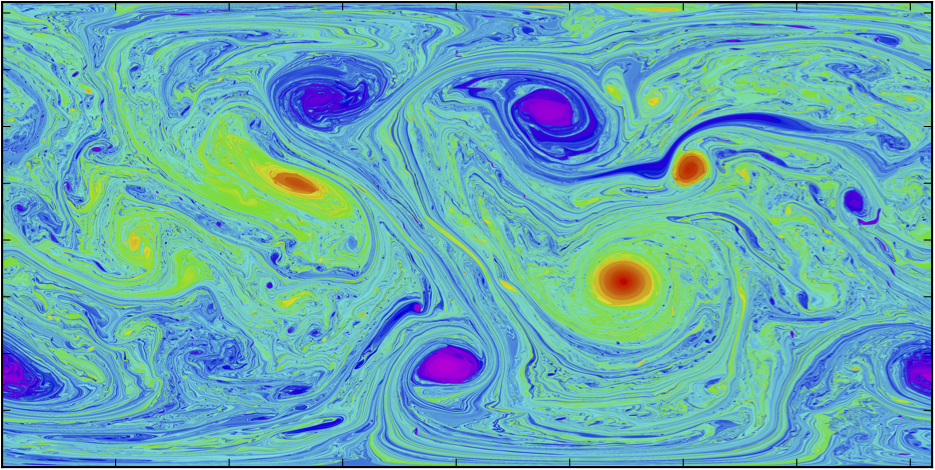}
\end{array}$

\end{center}

\caption{Early-intermediate time vorticity distribution
  $\omega(\hr,40)$, depicted as in figure~\ref{fig-vor4}, for (a) GGM
  at $D=655362$, (b) GGM at $D=2621442$, (c) CLAM at $n_g=128$, (d)
  CLAM at $n_g=256$, (e) CLAM at $n_g=512$ and (f) CLAM at
  $n_g=1024$. The percentage decay in enstrophy $Z$ by this time is
  56.8\%, 52.2\%, 
  50.0\%, 44.1\%, 38.3\% and 32.4\%, respectively.  
  In CLAM, the results are plotted on a grid 16 times finer in each
  direction except in (e) where every other fine grid point is plotted,
  and in (f) where every 4th grid point is plotted. 
  Same color scale as in Fig. \ref{fig-vor0}.}
\label{fig-vor40}
\end{figure}

By $t=40$, ten times further into the evolution, the simulations begin
to diverge --- see figure~\ref{fig-vor40} which includes now the two
highest resolution CLAM simulations at $n_g=512$ and $1024$.  By this
time, significant enstrophy dissipation has occurred, albeit least in
the highest resolution case.  Vorticity contours are stretched nearly
exponentially (at a rate comparable to $|\omega|_{\mathsf{max}}$, see
e.g.\ \cite{dritschel07}).  By incompressibility, this forces
filaments to thin exponentially and gradients to grow exponentially.
This behaviour cannot be followed indefinitely in numerical
simulations at finite resolution, so inevitably dissipation occurs.
In CLAM, unlike in GGM and other grid-based numerical
methods, the steep gradients can be preserved, without
dissipation, due to the contour representation of $\omega$ used in
CLAM.  This preservation of steep gradients is the key feature which
enables CLAM to model flows at very long times albeit at the cost of representing 
the continuous vorticity field by a discrete set of levels.  There is no
progressive erosion of vorticity gradients, as in grid-based methods.
The vorticity filaments, on the other hand, play a relatively benign
role in the dynamics.  They contribute negligibly to the flow field
and have little chance of rolling up into coherent vortices 
\citep{dritschel91}.

\begin{figure}
\begin{center}

$\begin{array}{c@{\hspace{0.12in}}c}
\multicolumn{1}{l}{\mbox{(a)}} &
\multicolumn{1}{l}{\mbox{(b)}}\\ [-0.0in]
\includegraphics[width = 2.56in]
{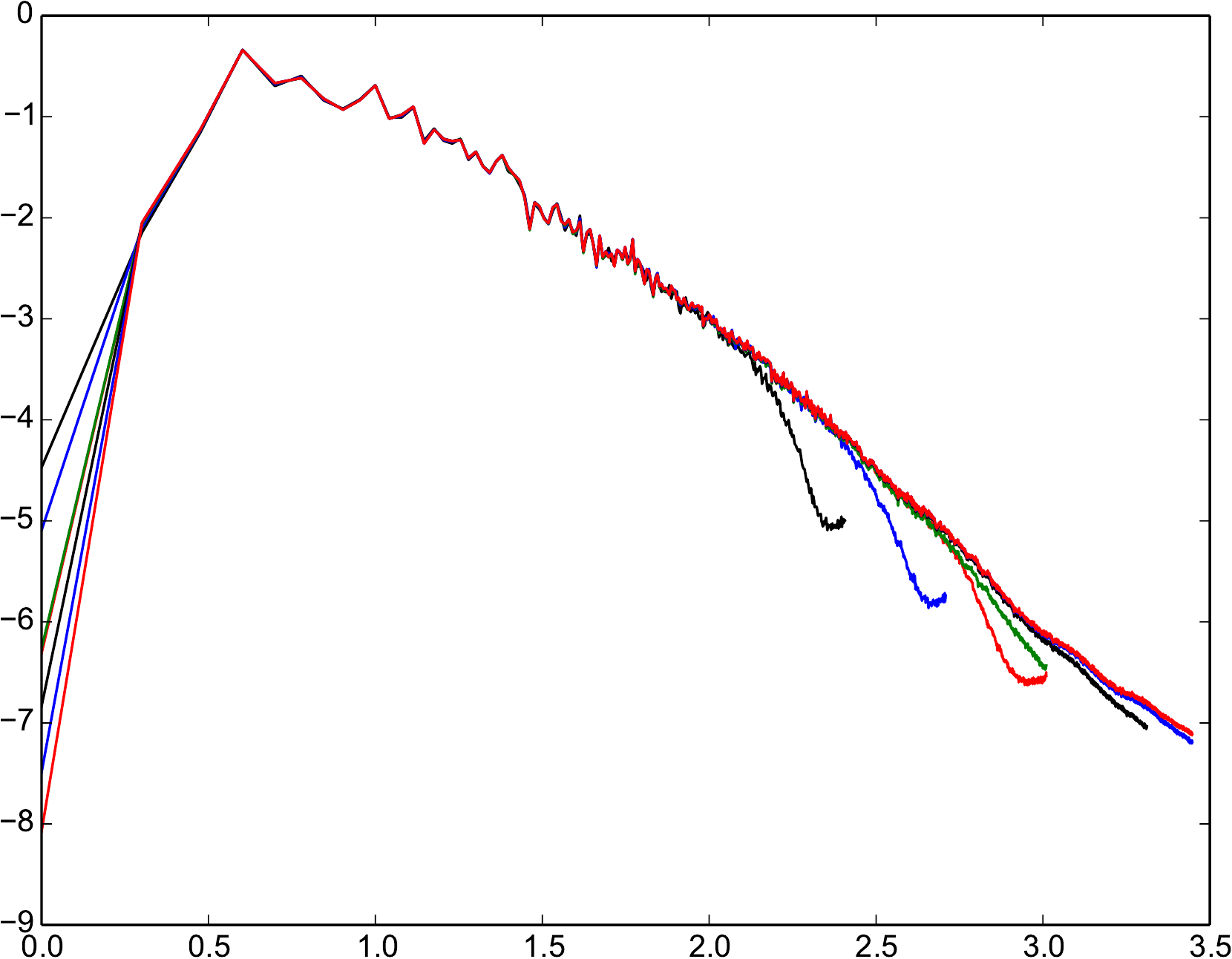} &
\includegraphics[width = 2.56in]
{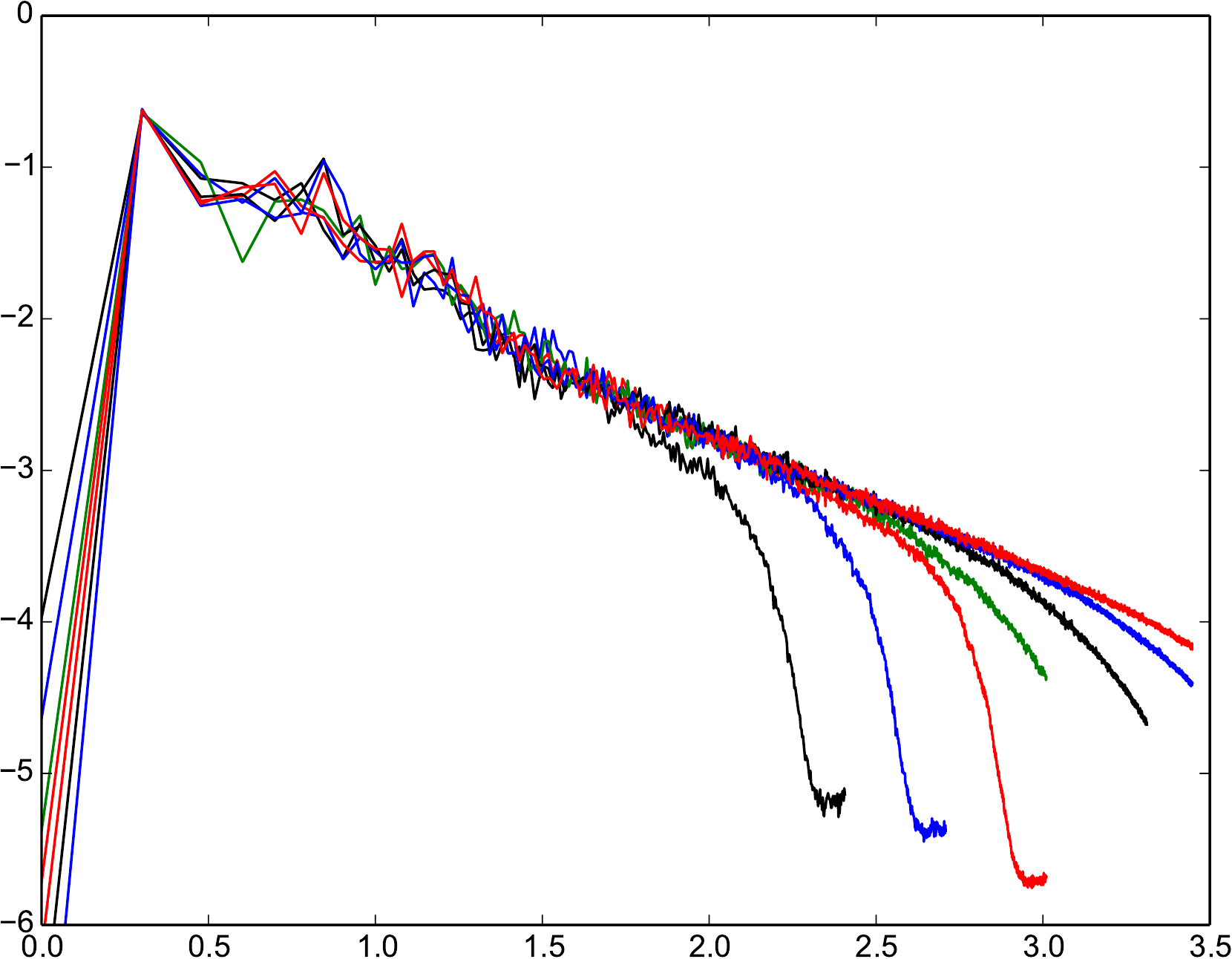}
\end{array}$

\end{center}
\caption{1D vorticity power spectra ${\mathcal{Z}}(l,t)$ at (a) at
  $t=4$ and (b) $t=40$.  Colours (online) cycle from black, to blue,
  to red, to green and back to black, starting from the lowest
  resolution GGM simulation ($D=163842$) and ending with the highest
  resolution CLAM one ($n_g=1024$).  Three spectra are shown for GGM
  and four are shown for CLAM.  Log (base 10) scales are used.}
\label{fig-earlyspectra}
\end{figure}

Notably, the CLAM simulations exhibit more structure than the GGM
ones.
The GGM simulations appear to be at lower
resolution, and the effects of numerical diffusion are 
visible.  The 1D vorticity power spectra ${\mathcal{Z}}(l,t)$, shown
in figure~\ref{fig-earlyspectra} at time $t=4$ and $40$, confirm this
impression.  The CLAM spectra are undissipated out to the maximum
wavenumber computed (which is limited to spherical wavenumber $\ell=2800$ in the SHTOOLS
software package used).  The GGM spectra turn down from hyperdiffusion
but also exhibit a rise near the maximum wavenumber (the
spectral amplitudes are however too weak to play a significant role).
Note the 
significant
shallowing of the spectrum in time, as nonlinearity transfers
a large fraction of the enstrophy to small scales.  The spectral slope
over the range $l \geq 100$ shallows from approximately $-2.9$ to
$-1.9$ between $t=4$ and $40$ in the highest resolution CLAM
simulations.

\begin{figure}
\begin{center}

$\begin{array}{c@{\hspace{0.12in}}c}
\multicolumn{1}{l}{\mbox{(a)}} &
\multicolumn{1}{l}{\mbox{(b)}}\\ [-0.0in]
\includegraphics[width = 2.56in]
{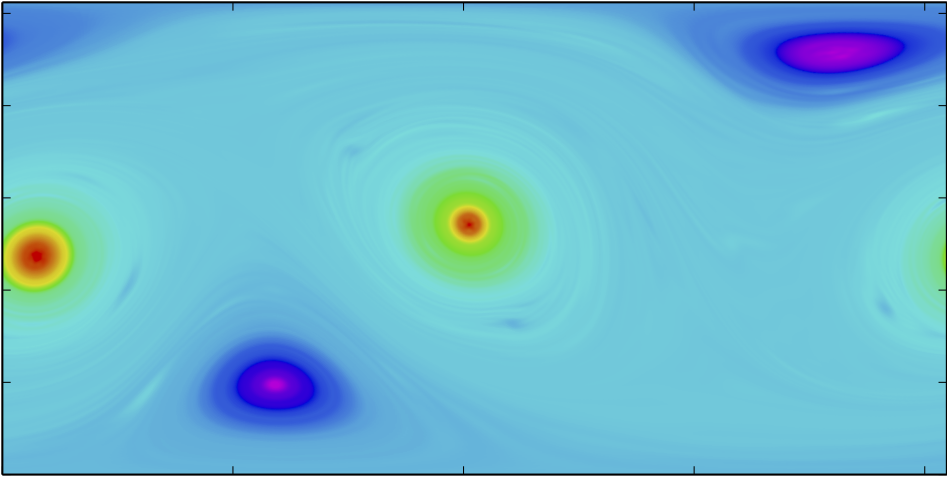} &
\includegraphics[width = 2.56in]
{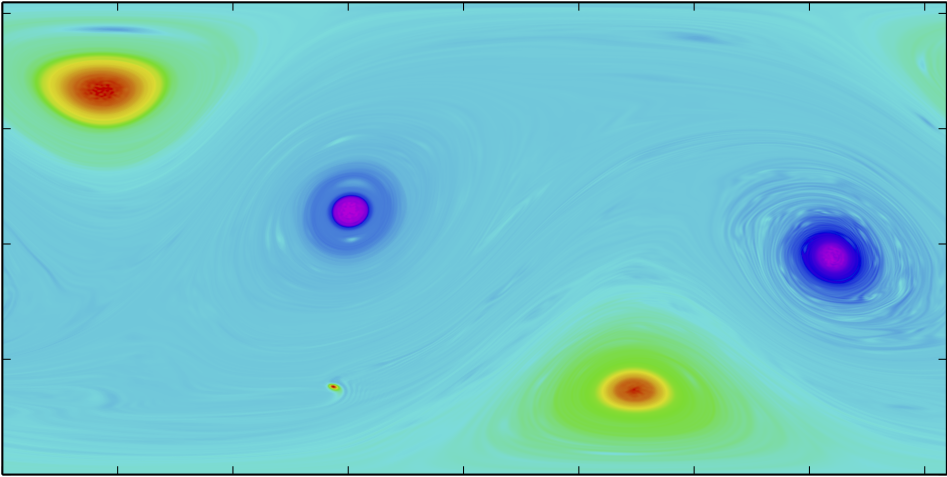}
\end{array}$

\vspace{0.1cm}

$\begin{array}{c@{\hspace{0.12in}}c}
\multicolumn{1}{l}{\mbox{(c)}} &
\multicolumn{1}{l}{\mbox{(d)}}\\ [-0.0in]
\includegraphics[width = 2.56in]
{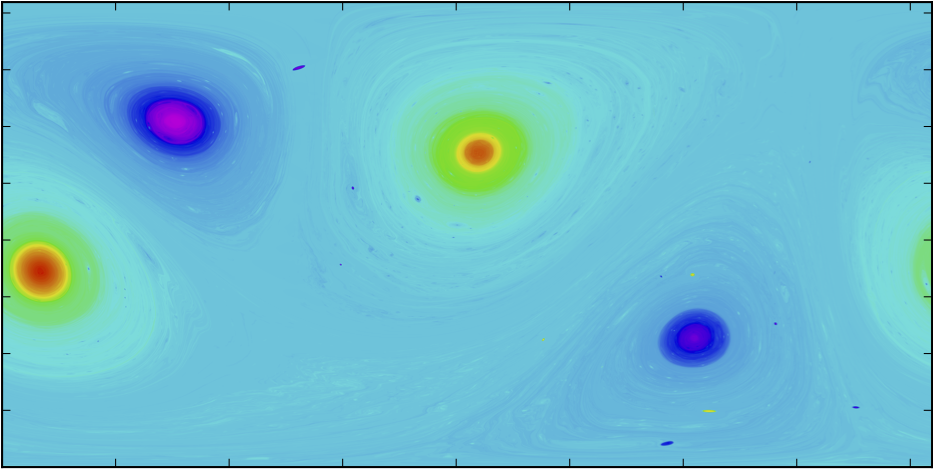} &
\includegraphics[width = 2.56in]
{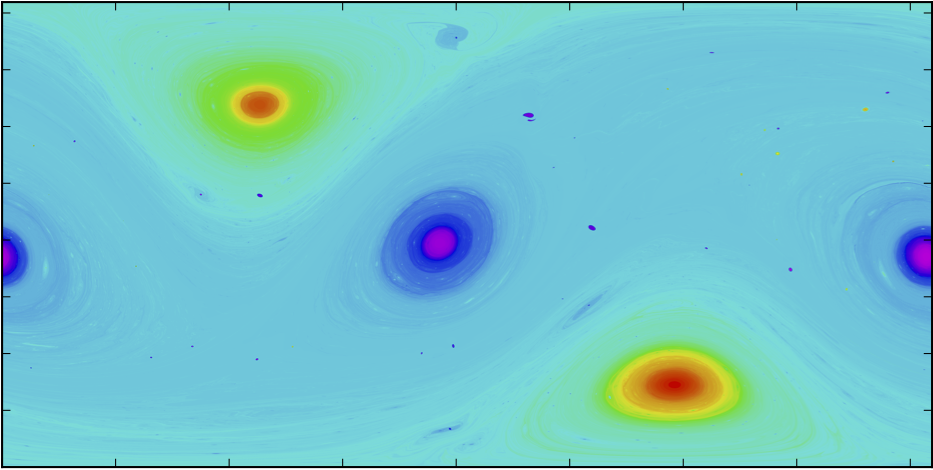}
\end{array}$

\end{center}

\caption{Intermediate-late time vorticity distribution
  $\omega(\hr,400)$, depicted as in figure~\ref{fig-vor4} \&
  \ref{fig-vor40}, for (a) GGM at $D=655362$, (b) GGM at $D=2621442$,
  (c) CLAM at $n_g=512$ and (d) CLAM at $n_g=1024$.  Note, only the
  two highest CLAM results are shown, but the lower resolution results
  are qualitatively similar.  By this time, the enstrophy $Z$ has
  decayed to approximately $21\%$ of its initial value in all cases.
  No further significant decay occurs to $t=4000$ ($<0.05\%$).  Same color scale as in Fig. \ref{fig-vor0}.}
\label{fig-vor400}
\end{figure}

\begin{figure}
\begin{center}

$\begin{array}{c@{\hspace{0.12in}}c}
\multicolumn{1}{l}{\mbox{(a)}} &
\multicolumn{1}{l}{\mbox{(b)}}\\ [-0.0in]
\includegraphics[width = 2.56in]
{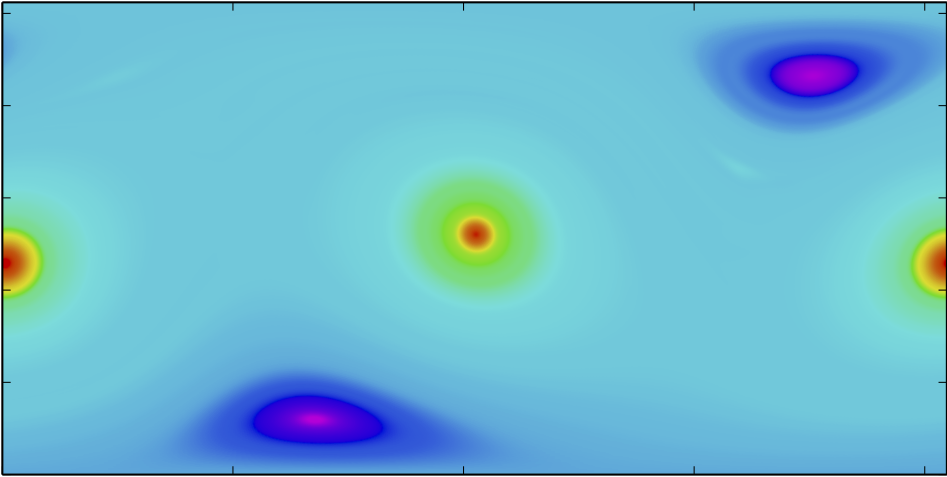} &
\includegraphics[width = 2.56in]
{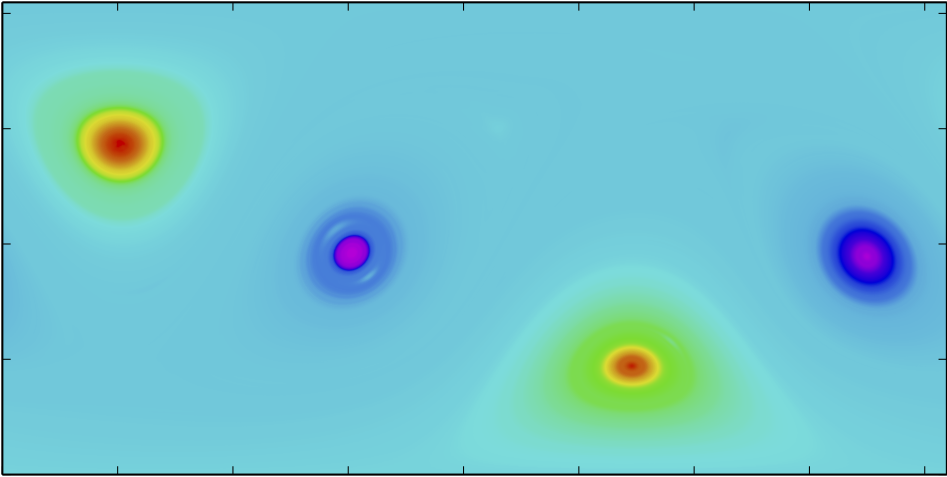}
\end{array}$

\vspace{0.1cm}

$\begin{array}{c@{\hspace{0.12in}}c}
\multicolumn{1}{l}{\mbox{(c)}} &
\multicolumn{1}{l}{\mbox{(d)}}\\ [-0.0in]
\includegraphics[width = 2.56in]
{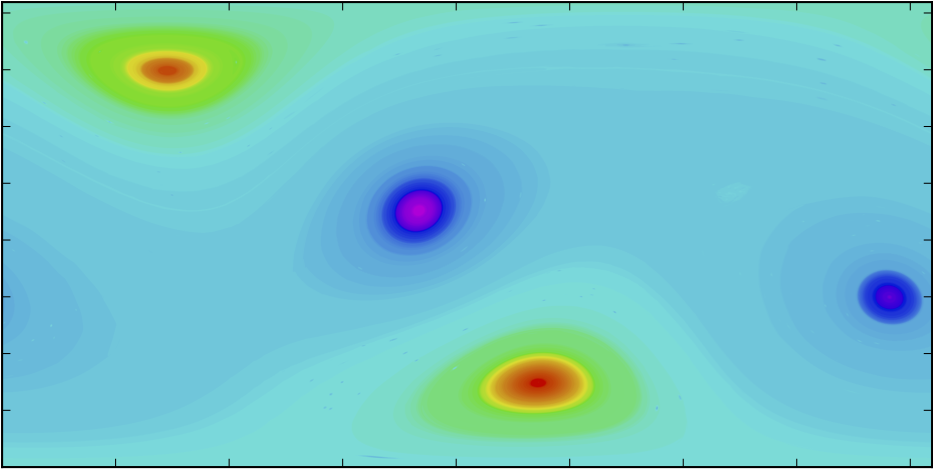} &
\includegraphics[width = 2.56in]
{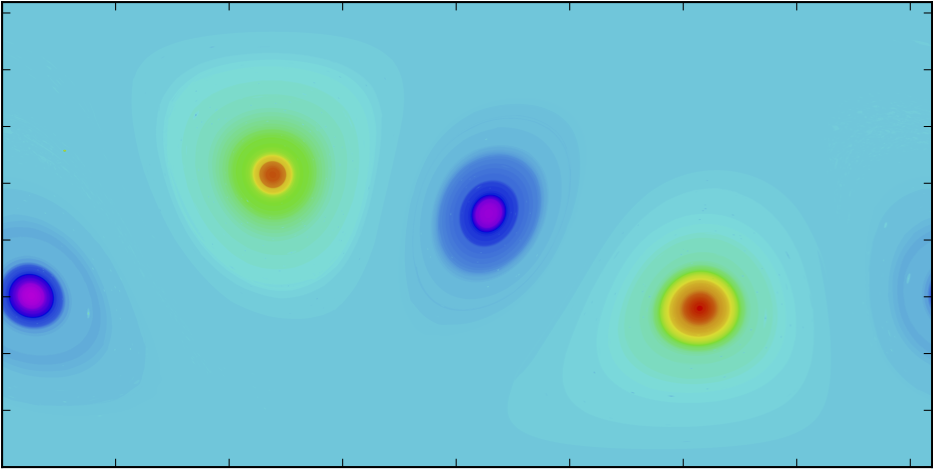}
\end{array}$

\end{center}

\caption{Late (and final) time vorticity distribution
  $\omega(\hr,4000)$, depicted as in figure~\ref{fig-vor4} \&
  \ref{fig-vor40}, for (a) GGM at $D=655362$, (b) GGM at $D=2621442$,
  (c) CLAM at $n_g=512$ and (d) CLAM at $n_g=1024$.  Same color scale as in Fig. \ref{fig-vor0}.}
\label{fig-vor4000}
\end{figure}

\begin{figure}
\begin{center}
\includegraphics[width = 5.24in]{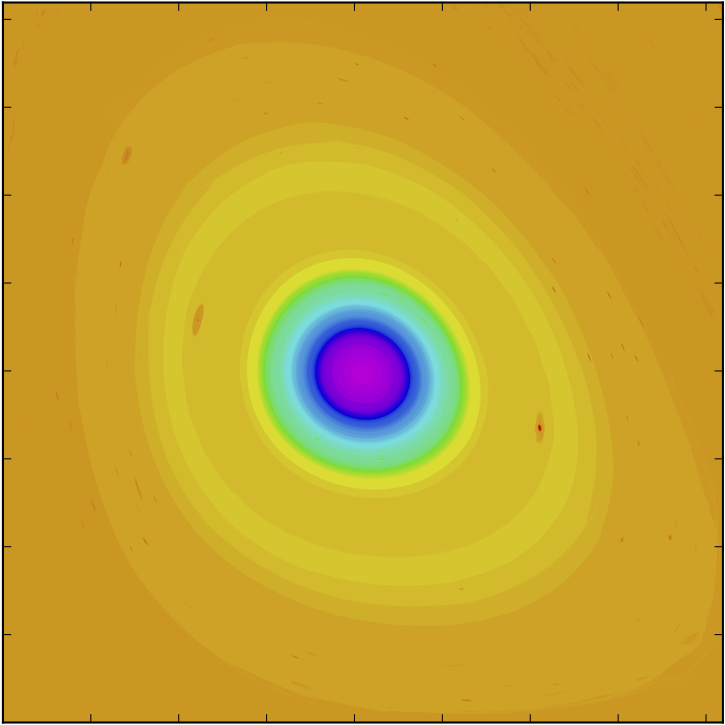}
\end{center}

\caption{Zoomed, full-resolution view at $t=4000$ of the vortex seen
  in the lower-left part of figure~\ref{fig-vor4000}(d).  The view is
  centred at longitude $-15\pi/16$ and latitude $-\pi/8$, and shows a
  $\pi/2$ range in longitudes and latitudes (1/8th of the domain or
  $8192^2$ grid points).  Colours (online) range from the minimum to
  the maximum vorticity in the domain of view.  Concentric steps seen in the vorticity are
  an artifact of the discretization of the vorticity levels in the CLAM algorithm.}
\label{fig-zoom4000}
\end{figure}

By $t=400$, a further ten times later in the evolution, much of the
complex structure evident at $t=40$ has cascaded beyond the maximum
scale resolved in all simulations --- see figure~\ref{fig-vor400} for
the two highest resolution GGM and CLAM results.  Nonetheless,
extensive small-scale structure survives, particularly in the CLAM
results.  The main quadrupolar vortices exhibit stepped distributions,
`vorticity staircases' (evident also in the GGM results), together
with a multitude of small-scale vortices caught between the sharp
vorticity gradients.  This structure persists, albeit somewhat
reduced, all the way to the end of the simulations at $t=4000$ (see
figure~\ref{fig-vor4000}).  A zoom at $t=4000$, shown in
figure~\ref{fig-zoom4000} for the highest resolution CLAM case,
reveals the extent of this fine-scale structure --- literally hundreds
of small-scale vortices exist between the vorticity steps on this
large-scale vortex alone.  Notably, virtually all of these small-scale
vortices have 
vorticity anomalies which cause the vortices to rotate with the shear 
associated with the differential rotation of the main large-scale vortex they
are embedded in.  This gives rise to stabilising `cooperative shear',
enabling them to persist for long times, perhaps indefinitely
\citep{dritschel90}.

The reduction in the size and number of small-scale vortices between
$t=400$ and $t=4000$ is likely to be a numerical artefact, the effect
of very weak dissipation accumulated over a very long time.  In CLAM,
every few eddy turnover times (about 7 or so units of time), contours
are rebuilt on an ultra-fine grid to prevent overlapping
\citep{fontane09,dritschel10}.  Each time this happens, inevitably
small contours tend to shrink slightly on average (and the effect is
strongest for the smallest resolved contours).  This, we believe, is
the primary cause for the reduction of small-scale structure.
Notably, each resolution doubling performed here reveals significantly
more structure persisting to the end of the simulation, and in the
limit of infinite resolution --- a perfect model --- we expect a
multitude of vortices occupying an ever extending range of scales.
Unlike in the case of inviscid 2D turbulence in an unbounded space
\citep{dritschel09a}, where vortices are free to roam and occasionally
collide to form both larger and smaller vortices, here the large-scale
flow imposed by the finite domain (the sphere) channels
vortices to move along the quasi-steady streamlines of each
large-scale vortex, preventing vortex collisions.  As a result,
vortices may persist indefinitely once the possibility of encountering
another vortex in a nearby streamline at close range has been
eliminated.

\begin{figure}
\begin{center}

$\begin{array}{c@{\hspace{0.12in}}c}
\multicolumn{1}{l}{\mbox{(a)}} &
\multicolumn{1}{l}{\mbox{(b)}}\\ [-0.0in]
\includegraphics[width = 2.56in]
{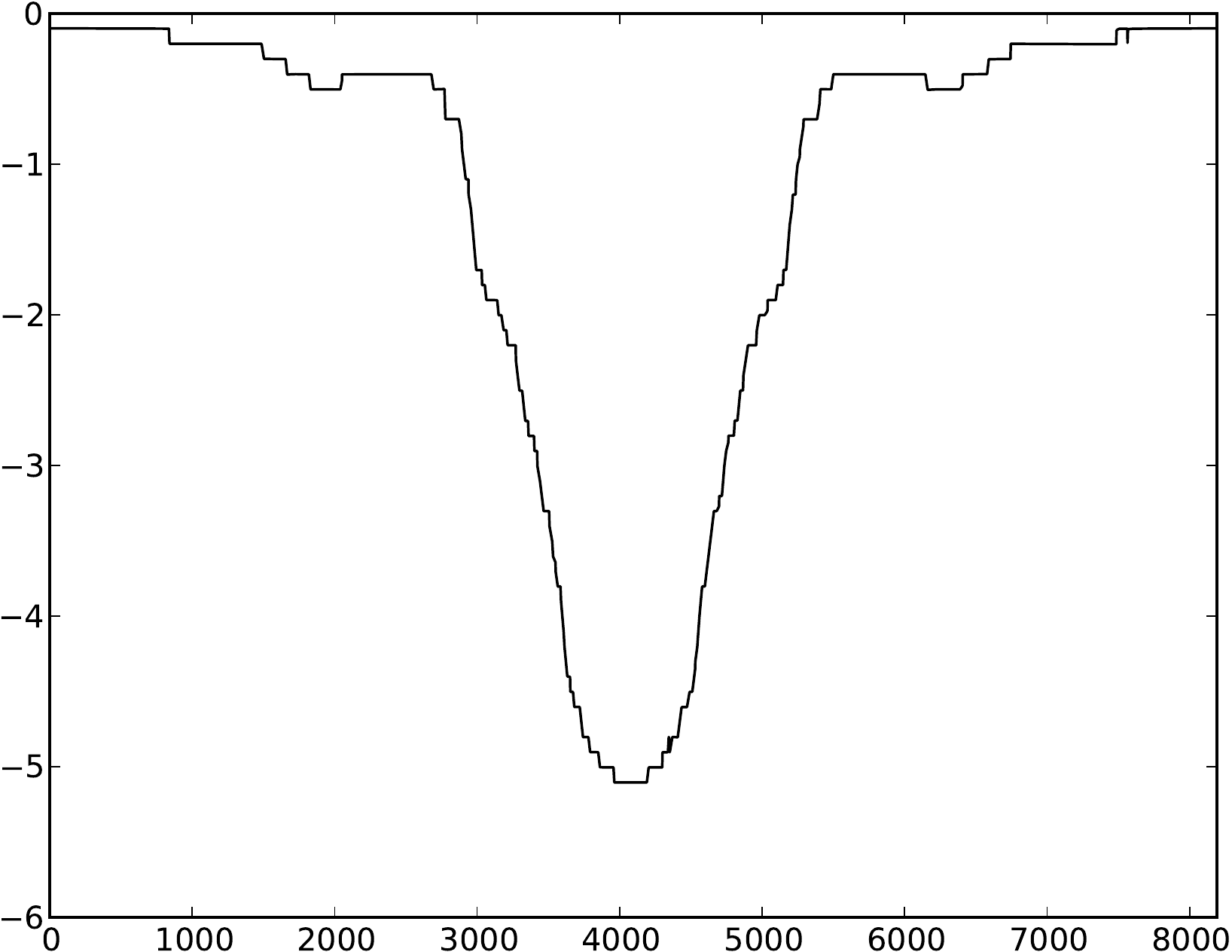} &
\includegraphics[width = 2.56in]
{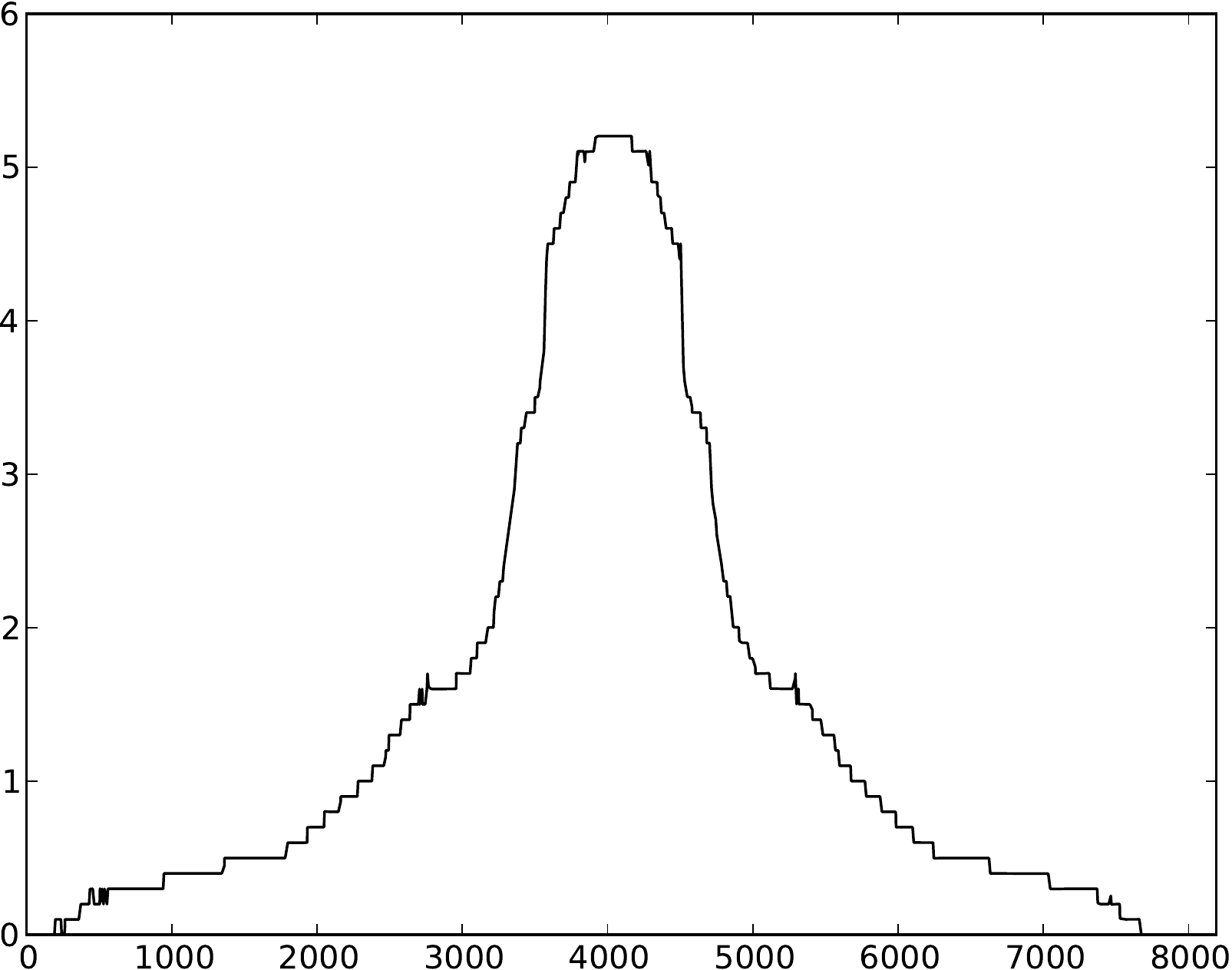}
\end{array}$

\vspace{0.1cm}

$\begin{array}{c@{\hspace{0.12in}}c}
\multicolumn{1}{l}{\mbox{(c)}} &
\multicolumn{1}{l}{\mbox{(d)}}\\ [-0.0in]
\includegraphics[width = 2.56in]
{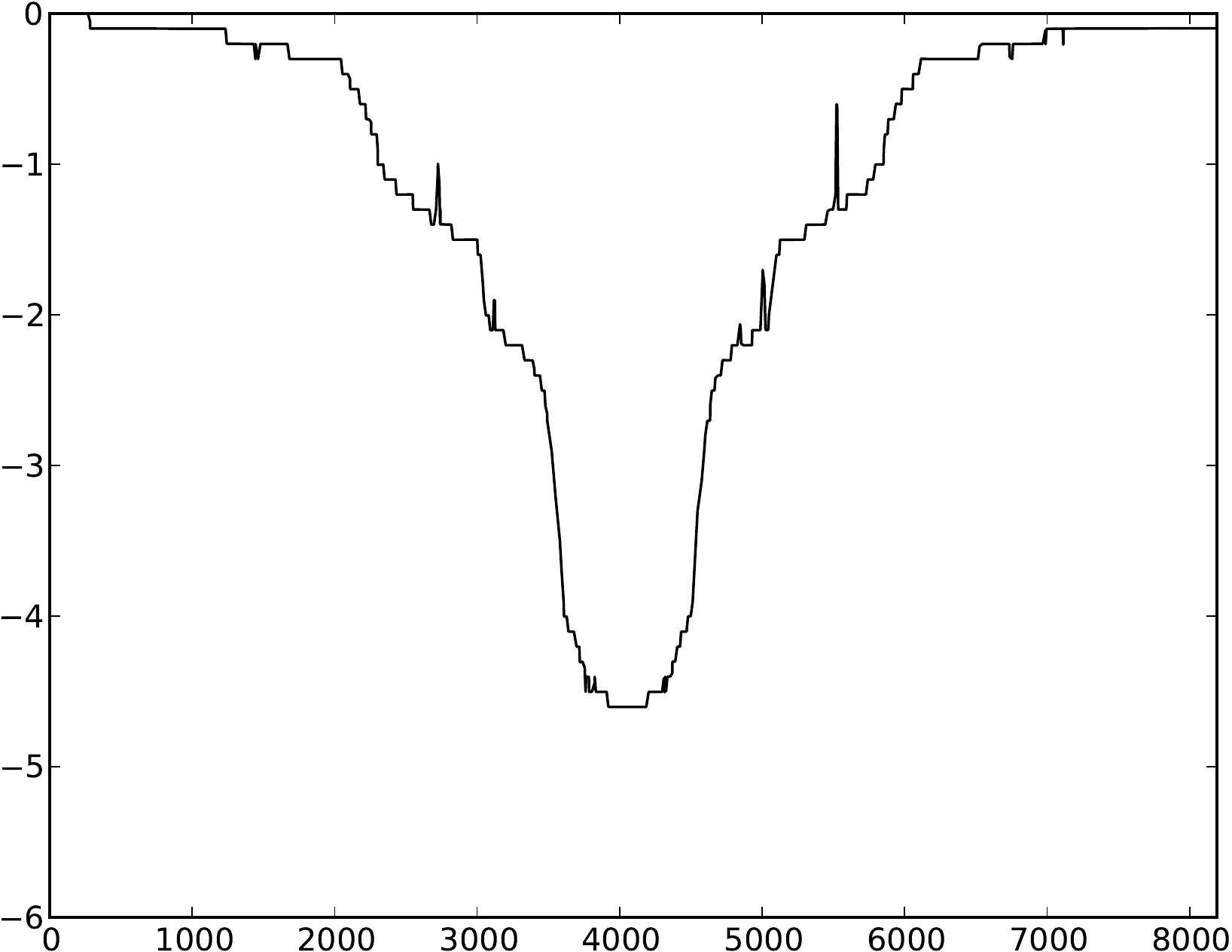} &
\includegraphics[width = 2.56in]
{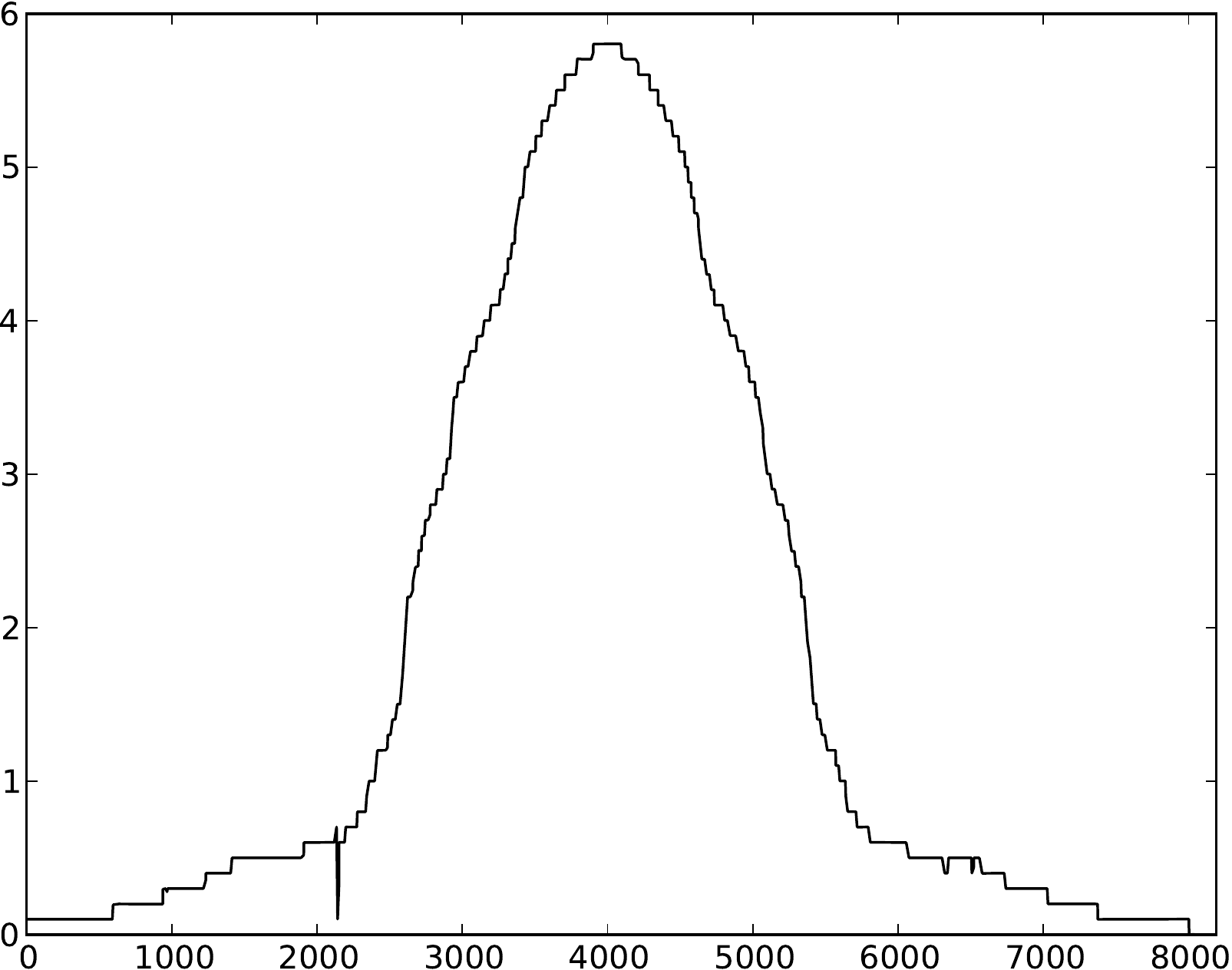}
\end{array}$

\end{center}

\caption{(a)--(d) Constant-latitude cross sections of the vorticity
  field taken through each of the 4 large-scale vortices (from left to
  right in figure~\ref{fig-vor4000}(d)) in the highest resolution CLAM
  simulation at $t=4000$. The relative longitudinal grid point is
  indicated on the horizontal axis. The small steps show the vorticity
  contour level spacing used in the CLAM simulation. The important
  features are the large steps, seen in many of the vortices. Note
  also that the cross sections occasionally pass through a small-scale
  vortex.}
\label{fig-cross4000}
\end{figure}

Vorticity cross-sections through the 4 main vortices in
figure~\ref{fig-vor4000}(d) exhibit stepped vorticity distributions,
with alternating steep and shallow gradients (see
figure~\ref{fig-cross4000}).  These are also found in the highest
resolution GGM simulation, together with small-scale vortices (though
many fewer due to the effects of hyperdiffusion).  The stepped
vorticity distributions seen here appear to be closely analogous to
those arising spontaneously in rotating 2D turbulent flows having a
mean `planetary' vorticity gradient (see \cite{dritschel08} for a
review).  There, turbulence cannot entirely break down this gradient
but rather mixes vorticity inhomogenously, resulting in a stepped
vorticity distribution in the inviscid limit \citep{scott12}.  The
associated flow consists of a series of eastward (prograde) jets
centred on each step, reminiscent of the jets observed in many
planetary atmospheres, most notably Jupiter's.  In the simulations
conducted here, each large-scale vortex has a vorticity gradient which
is reshaped into a stepped distribution early on in the turbulent flow
evolution.

\begin{figure}
\begin{center}

$\begin{array}{c@{\hspace{0.12in}}c}
\multicolumn{1}{l}{\mbox{(a)}} &
\multicolumn{1}{l}{\mbox{(b)}}\\ [-0.0in]
\includegraphics[width = 2.56in]
{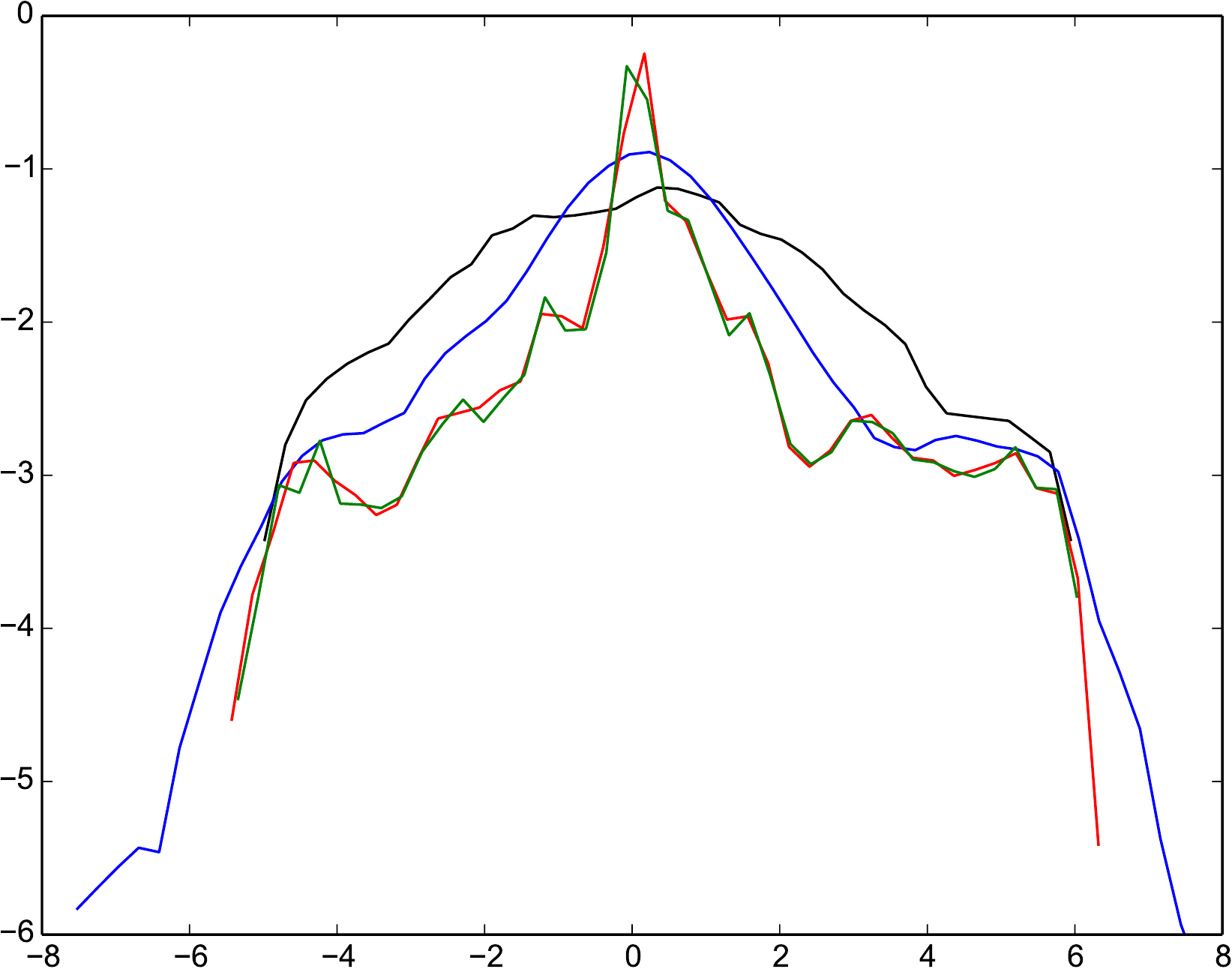} &
\includegraphics[width = 2.56in]
{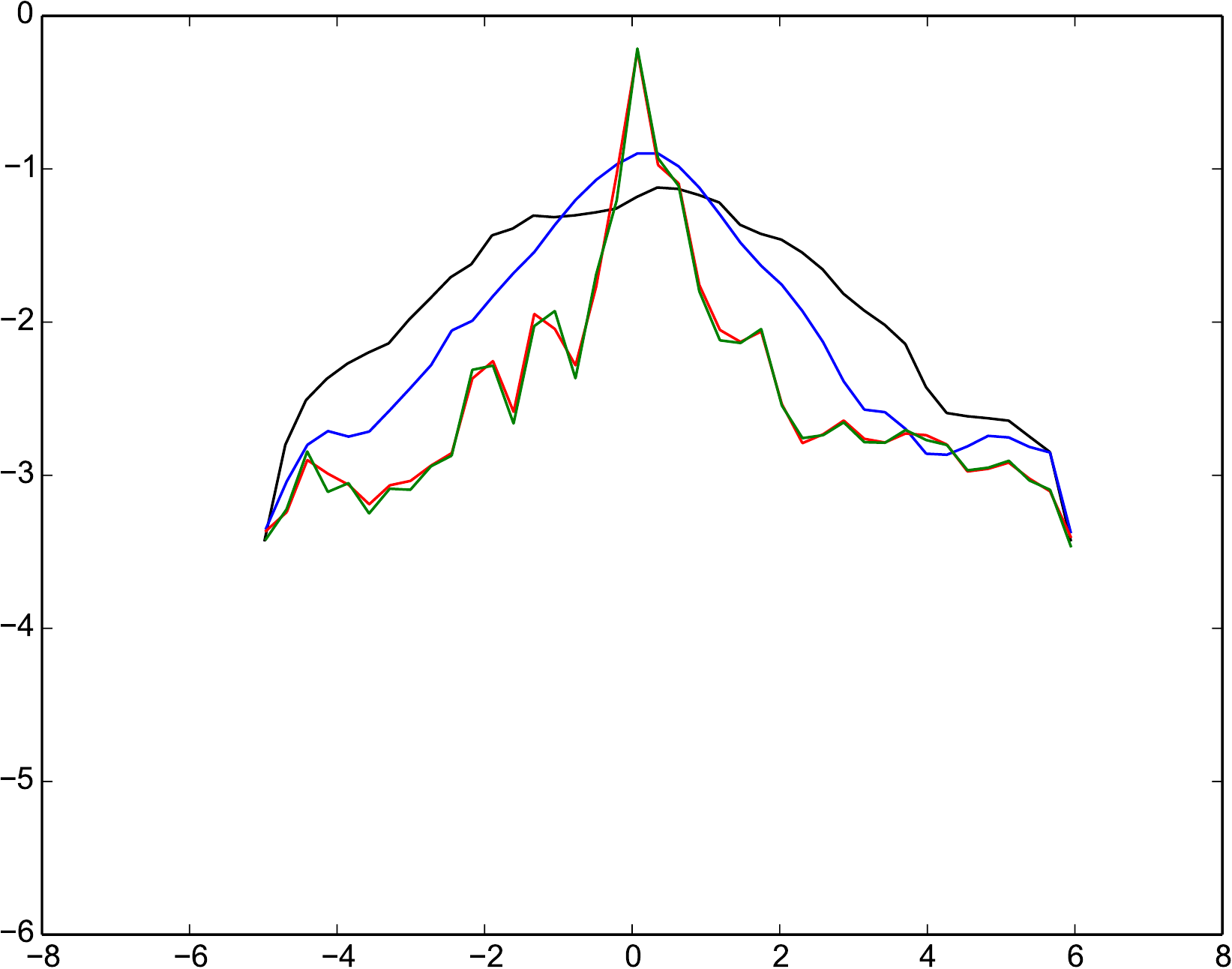}
\end{array}$

\end{center}

\caption{The log-scaled vorticity measure $\log_{10}p(\omega,t)$ at
  selected times in (a) the highest resolution GGM simulation, and (b)
  the highest resolution CLAM simulation.  The times shown (colours
  online) are $t=0$ (black), $t=40$ (blue), $t=400$ (red) and $t=4000$
  (green).}
\label{fig-pdfevol}
\end{figure}

The impact of this turbulent mixing can also be seen in the area
fraction occupied by each vorticity level --- the so-called `vorticity
measure' $p(\omega,t)$.  This is shown at a few selected times in
figure~\ref{fig-pdfevol} for the highest resolution GGM and CLAM
simulations, in (a) and (b) respectively.  In a perfectly inviscid
fluid, $p(\omega,t)$ is time-independent and generates the conserved
Casimirs in (\ref{casimirs}) through the relations
$\cC_n=4\pi\int\omega^n p(\omega)\der\omega$, $n=2,\,3,\,...$.  The
lack of conservation seen here is caused by the inevitable cascade of
vorticity to small scales --- none of the Casimirs can be preserved in
simulations at finite resolution (indeed, $\cC_2$ decays by $79\%$
here).  The initial vorticity measure $p(\omega,0)$ is broadly
distributed between $\omega_{\mathsf{min}}=-5.1187$ and
$\omega_{\mathsf{max}}=5.7996$, then narrows and diminishes nearly
everywhere except for values of $\omega$ near $0$.  By $t=400$, the
vorticity measure has converged to a nearly fixed form, which is
noticeably jagged, perhaps related to the non-monotonic vorticity
profiles seen in figure~\ref{fig-cross4000}(a).  
In the CLAM results on
the right, notice that $\omega_{\mathsf{min}}$ and
$\omega_{\mathsf{max}}$ are conserved, and that the area occupied by
these extreme vorticity values does not change significantly over
time.  This is sensible physically, as the extreme values are most
resilient to deformation.  On the other hand, in the GGM results,
there is clear evidence of hyperviscous overshoot: the extreme values
are not conserved (at times not shown here they are nearly twice their
initial values).  Slightly greater and unphysical extreme values
persist to the end of the simulation, though they have very small measure.  
In both simulations, the
observed strong growth of $p(\omega,t)$ near $\omega=0$ is due to the
efficient mixing of weak vorticity, which essentially behaves as a
passive scalar.  As $|\omega|$ increases, it becomes increasingly
difficult to shear out vorticity structures and mixing becomes less
efficient.

\begin{figure}
\begin{center}

$\begin{array}{c@{\hspace{0.12in}}c}
\multicolumn{1}{l}{\mbox{(a)}} &
\multicolumn{1}{l}{\mbox{(b)}}\\ [-0.0in]
\includegraphics[width = 2.56in]{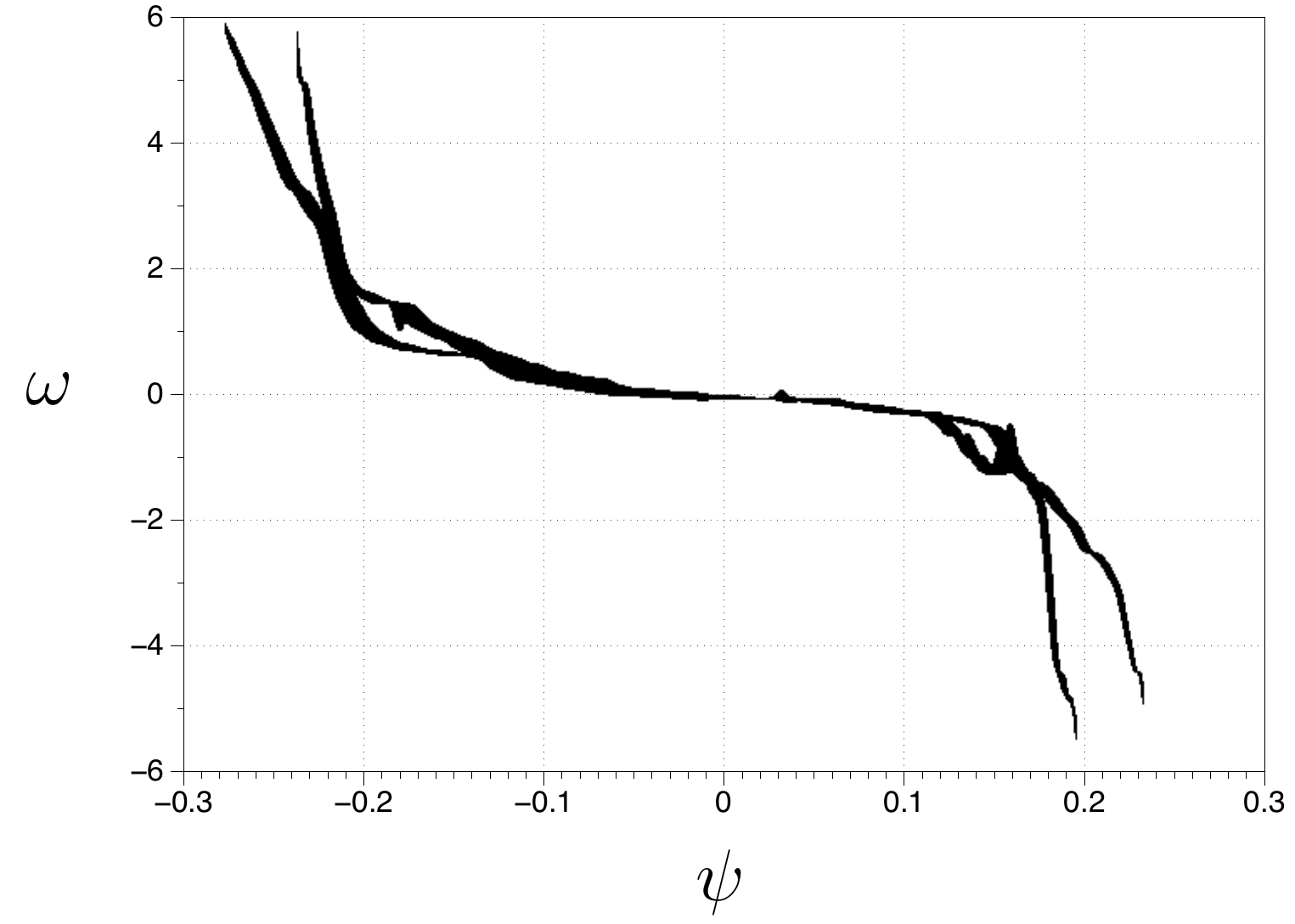} &
\includegraphics[width = 2.56in]{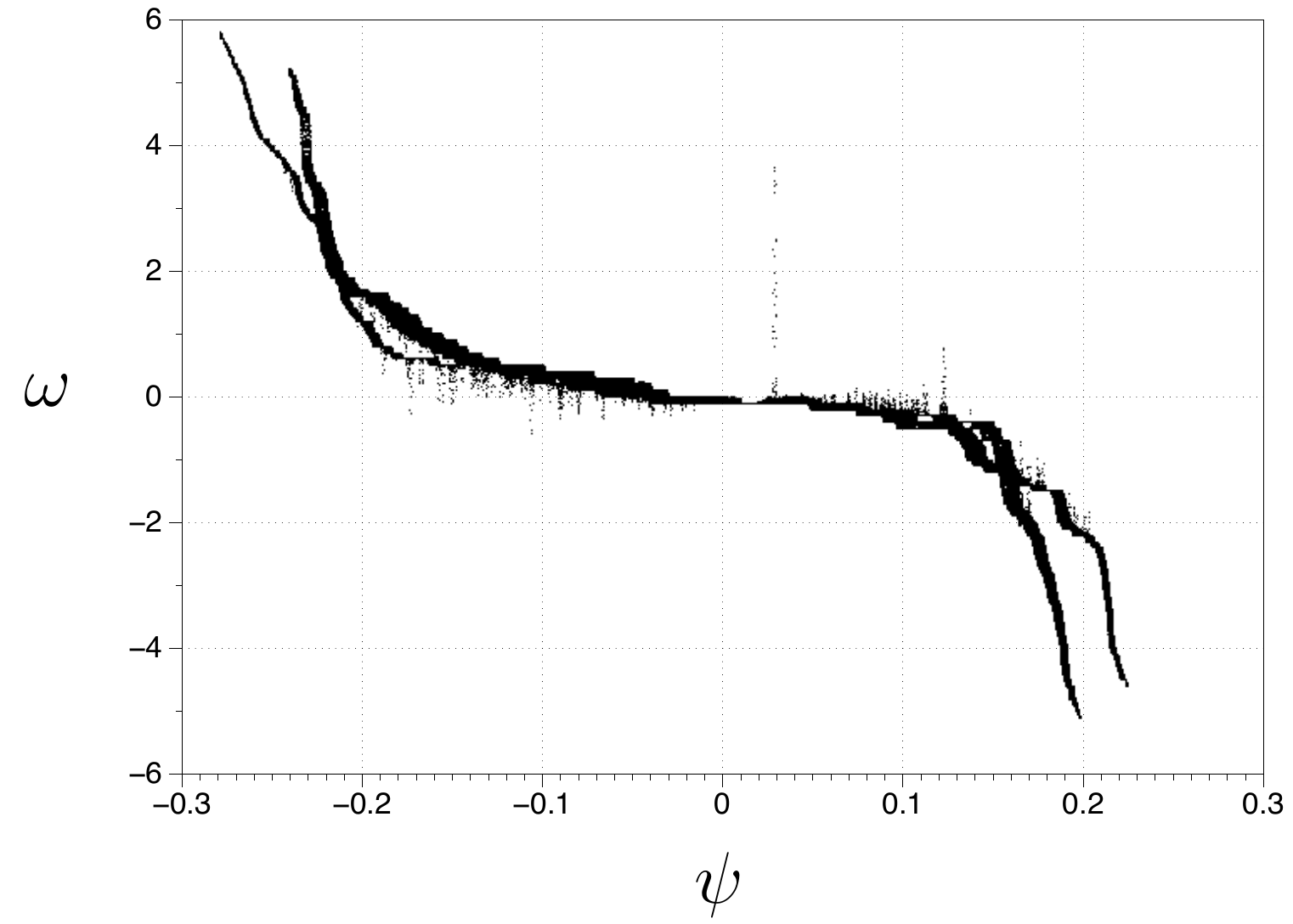}
\end{array}$

\end{center}

\caption{Scatter plots of vorticity $\omega$ vs. streamfunction $\psi$ 
at $t=4000$ in (a) the highest resolution GGM simulation, and (b)
  the highest resolution CLAM simulation.}
\label{fig-scatter4000}
\end{figure}

\subsection {Unsteadiness}
\label{sec:unsteady}

We next address the possible long-term equilibration of the flow, a
core prediction of the MRS statistical-mechanical theory.  As $t \to
\infty$, that theory predicts a steady flow characterised by a
functional relation between vorticity and streamfunction
$\omega=F(\psi)$.  The precise functional $F$ depends on the energy,
angular momentum (here zero), circulation (here zero due to the absence of boundary) and the unobservable higher-order fine-grained Casimirs. The resolved `coarse-grained' Casimirs
associated with the non-cascading vorticity field are often used to approximately determine the values of the fine-grained Casimirs (see e.g.\
\cite{qi14b} \& references therein).  The details do not matter
for what concerns us here.  If the flow is steady, then $F$ can
be obtained from a scatter-plot of $\psi$ vs $\omega$, in which
$F$ would appear as a single curve.  Our results are shown for
both GGM and CLAM at the highest resolutions and at the final time
in figure~\ref{fig-scatter4000}.  First of all, there are clearly
multiple branches, each corresponding to one of the main vortices.
Second, the data do not collapse on a line, but rather in a band,
indicating that the flow is {\it not} steady (see below for further
evidence).  Third, in the CLAM results, one sees a myriad of small-scale
vortices, due to the much weaker level of dissipation present.  In fact,
due to the need to keep the figure size manageable, the CLAM results are
downsampled by a factor of 16 in each direction.  The true number of 
small-scale vortices is far greater.  Figure~\ref{fig-scatter4000}(b)
gives a greatly simplified view.  The upshot is that the flow remains
unsteady and extremely complex, even despite the inevitable 
dissipation in CLAM which acts to remove small-scale structure over
time (see discussion above).

\begin{figure}
\begin{center}

$\begin{array}{c@{\hspace{0.12in}}c}
\multicolumn{1}{l}{\mbox{(a)}} &
\multicolumn{1}{l}{\mbox{(b)}}\\ [-0.0in]
\includegraphics[width = 2.56in]
{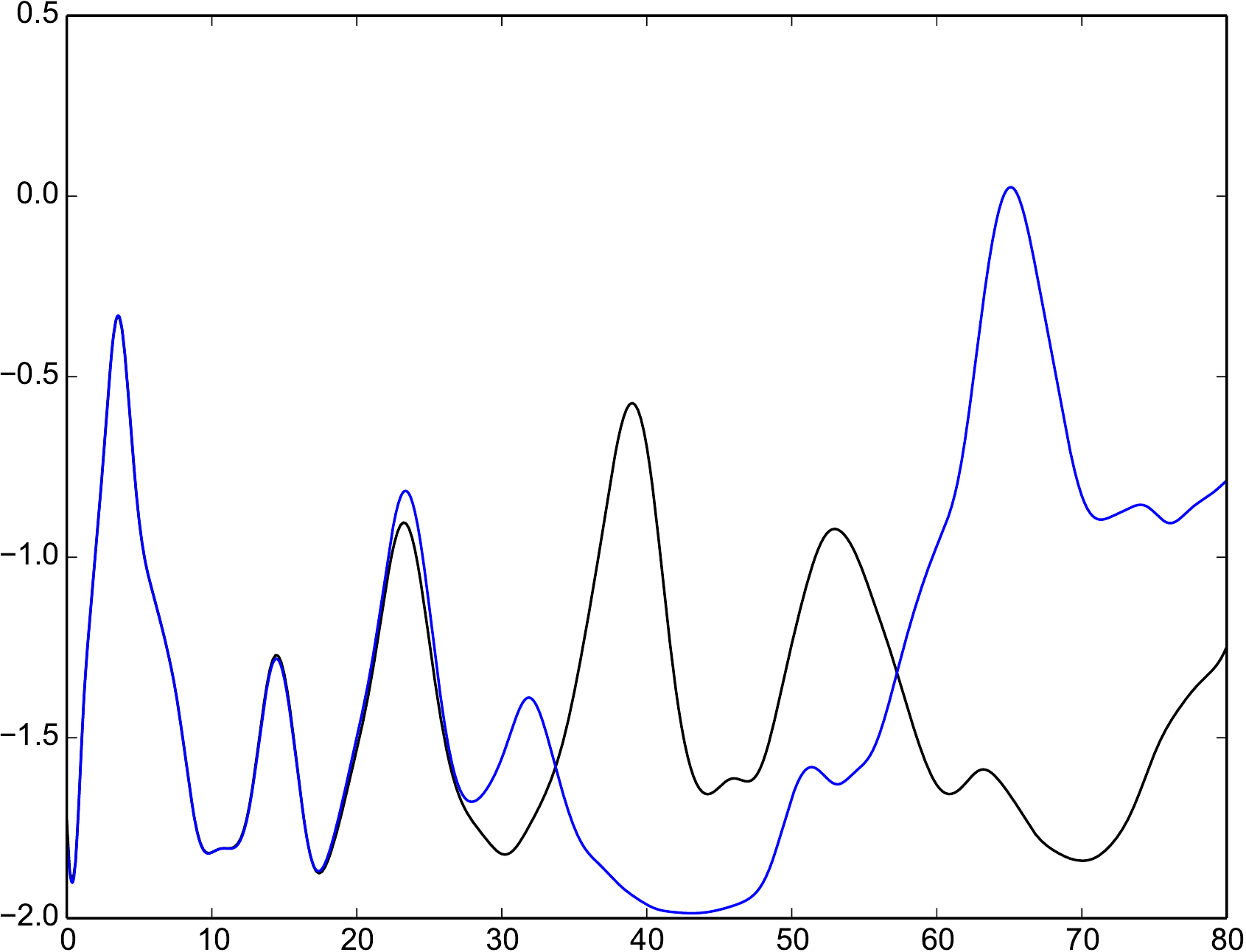} &
\includegraphics[width = 2.56in]
{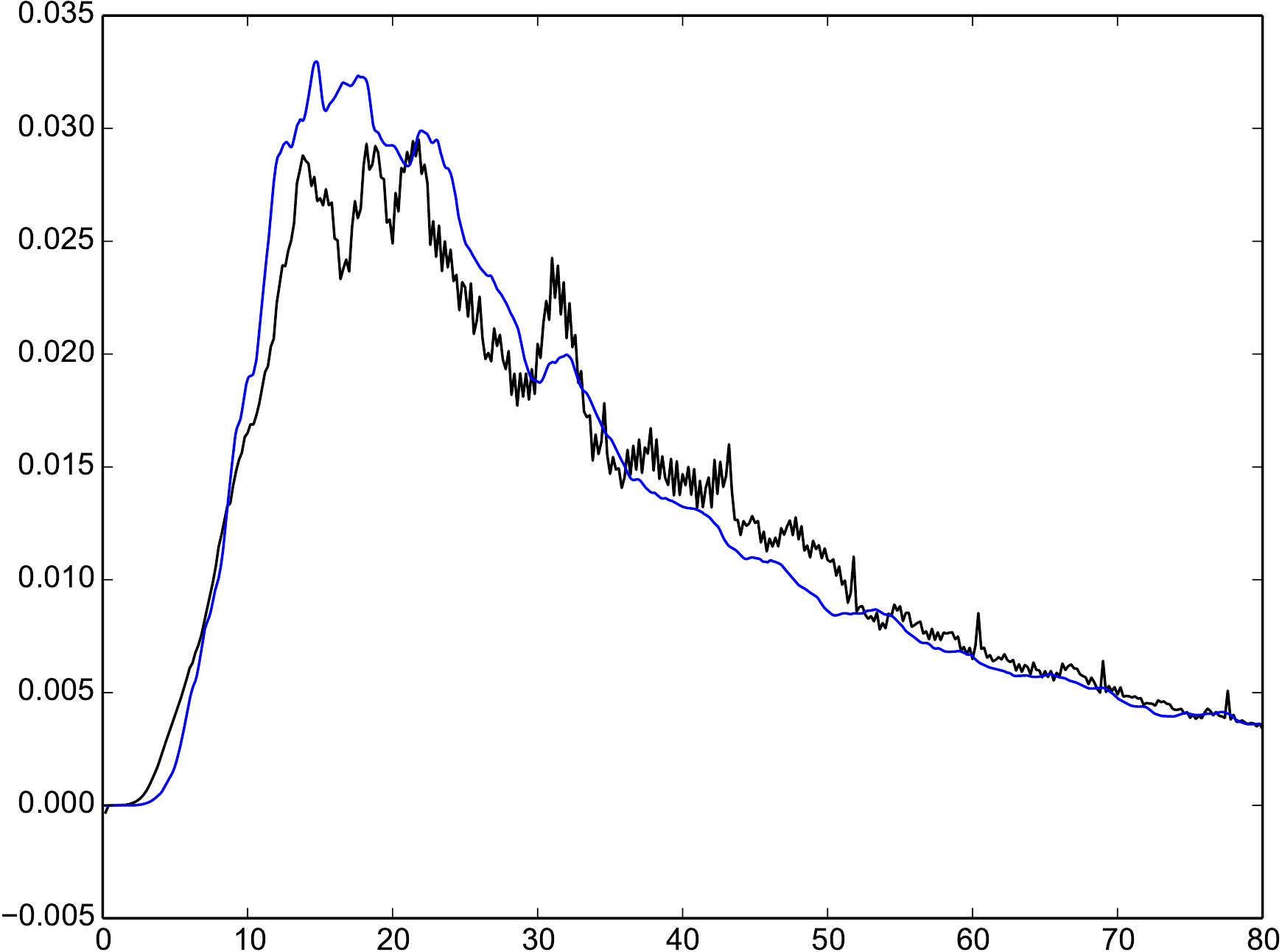}
\end{array}$

\end{center}
\caption{Early time evolution ($0\leq t \leq 80$) of (a) the
  `configuration' $W(t)$ (defined in the text) and (b) the 
  rate of enstrophy dissipation $-\der{Z}/\der{t}$, for both the
  highest-resolution CLAM and GGM simulations. Colours (online) show
  CLAM in black and GGM in blue.}
\label{fig-earlyW}
\end{figure}

An alternative way of measuring the unsteadiness of the flow was
introduced by \cite{qi14b}.  They developed a rotation and amplitude
independent measure of the flow configuration that involves only the 
streamfunction amplitudes $\psi_{\ell m}$ of the spherical harmonics $Y_\ell^m$ of degree
$\ell=2$ (these coefficients characterise the main quadrupolar
pattern seen at late times).  The order $m$ satisfies $-2 \leq m \leq 2$.  
Using the standard convention for spherical harmonics 
(for which $Y_\ell^{-m}=(-1)^m Y_\ell^{m*} $ where $*$ denotes complex conjugation), 
the configuration is obtained from the amplitudes: 
\begin{equation}
W(t)=\frac{-2~ \psi_{20}^3 + 6\psi_{20}~ (\psi_{2,-1}~ \psi_{21} + 2~ \psi_{2,-2}~ \psi_{22}) 
- 3\sqrt{6}~ (\psi_{2,-2}~ \psi_{21}^2 + \psi_{2,-1}^2~ \psi_{22})}
{\left(\psi_{2,-2}^2 + \psi_{2,-1}^2 + \psi_{20}^2 + \psi_{21}^2 + \psi_{22}^2\right)\sth} .
\label{config}
\end{equation}
$W$ ranges from $-2$ to $+2$; when $|W|=2$, there are just two
vortices, while when $W=0$ there are four of equal intensity.  The
sign of $W$ indicates the sign of the dominant pair of same-signed
vortices.

\begin{figure}
\begin{center}
\includegraphics[width = 5.24in]{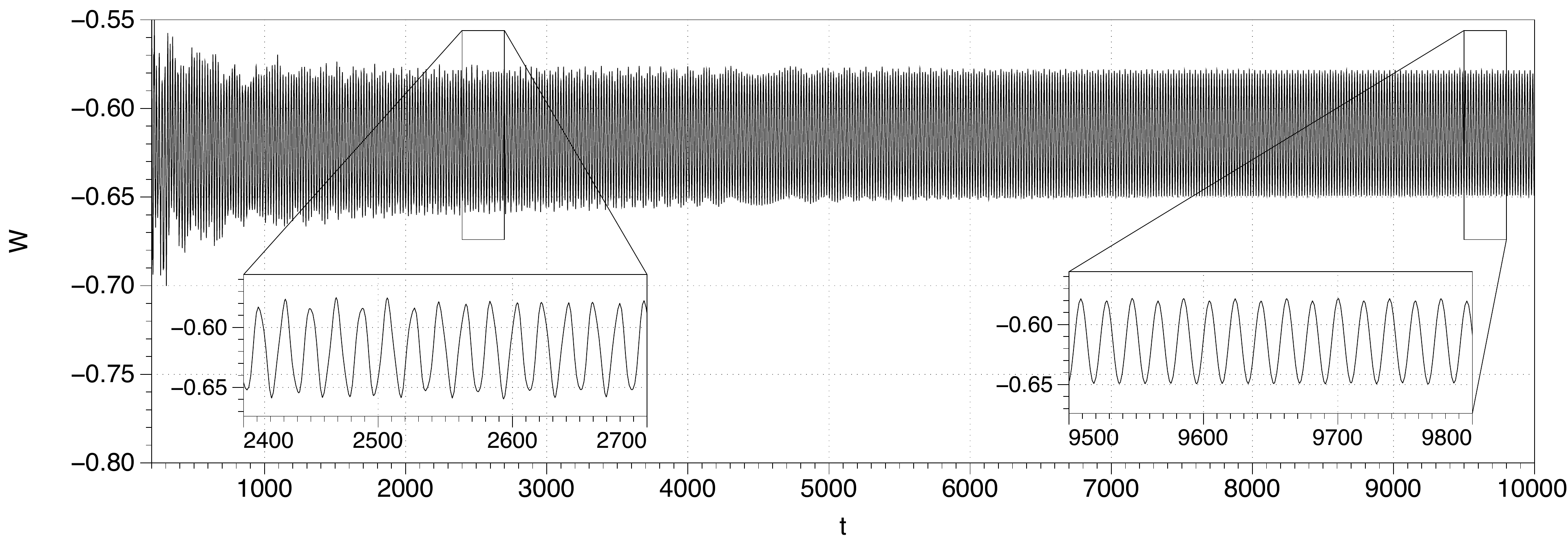} \\
\includegraphics[width = 5.24in]{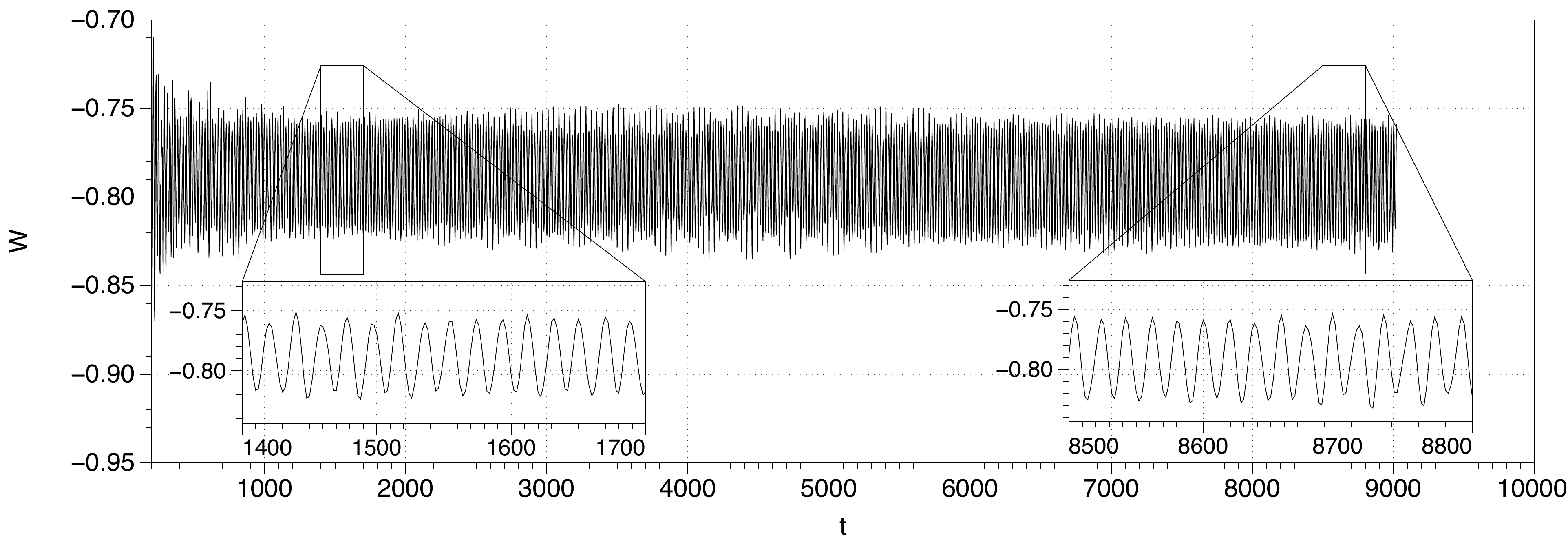}
\end{center}

\caption{Late time evolution ($t \geq 200$) of the `configuration' $W(t)$ 
  for both the second highest resolution GGM simulation (top) and highest 
  resolution CLAM simulation (bottom).}
\label{fig-lateW}
\end{figure}

The evolution $W(t)$ at early times, $0\leq t \leq 80$, in both GGM
and CLAM at the highest resolutions is compared in
figure~\ref{fig-earlyW}(a).  The correspondence is excellent until around
$t=20$, a little after the time of peak enstrophy dissipation rate
$-\der{Z}/\der{t}$ (shown in (b)).  After this
time, the results diverge due to the loss of predictability.  However,
both methods --- at all resolutions --- exhibit the same qualitative
behaviour in $W(t)$ thereafter: quasi-periodic and non-diminishing
oscillations, i.e.\ persistent unsteadiness.  This is shown in
figure~\ref{fig-lateW} for two very long integrations carried out to
around $t=10000$; the top panel shows GGM at $D=655362$ while the
bottom shown CLAM at $n_g=1024$ (both for $t \geq 200$).  Movies of the flow
evolution indicate that the oscillations are caused by the four main
vortices moving around each other, with global-scale excursions, and
with no sign of relaxing to equilibrium. Note that similar oscillations are also seen in the dipole case of figure 2 of \citet{morita11} where they are described as `small perturbations' without detailed discussion.

A simple model of this vortex motion is to consider just 4 point
vortices, singular distributions, in place of the continuous vorticity
distribution above.  For point vortices, the dynamical system has just
3 degrees of freedom, after taking into account conservation of
energy, angular momentum and the fact that the dynamics depends only
on the relative (angular) separation of the vortices
\citep{polvani93,newton01}.  The vortex motion is therefore either
quasi-periodic or chaotic in general; steady configurations are rare
(indeed of zero probability for arbitrary choices of the vortex
circulations and positions).

To test the effectiveness of this simple model, the 4 main vortices from
the highest resolution CLAM simulation at $t=4000$ (see 
figures~\ref{fig-vor4000} and \ref{fig-cross4000}) were identified by
locating contiguous vorticity regions having $|\omega| > 0.2$, a small
threshold value enabling one to capture most of the circulation within
each vortex.  The circulations $\Gamma_k$ ($k=1,\,2,\,3$ \& $4$) were 
found by area integration over each contiguous region of vorticity $R_k$, 
and the centres $\hr_k$ from 
\begin{equation}
\hr_k = \frac{1}{\Gamma_k}\int\!\!\int_{R_k} \omega\hr\dom .
\label{centres}
\end{equation}
This procedure however does not lead to zero angular impulse $\LL$, which
for point vortices is equal to $\sum_k \Gamma_k \hr_k$.  Since $\LL=0$
in the original flow, we adjust the circulations $\Gamma_k$ by subtracting
$\blam\bcdot\hr_k$ for each k, where $\blam$ is determined from a 
$3 \times 3$ linear system.  This adjustment is not biased toward any
one vortex and leads to a 10 to 15\% change in the circulations.  

\begin{figure}
\begin{center}
\includegraphics[width = 2.56in]{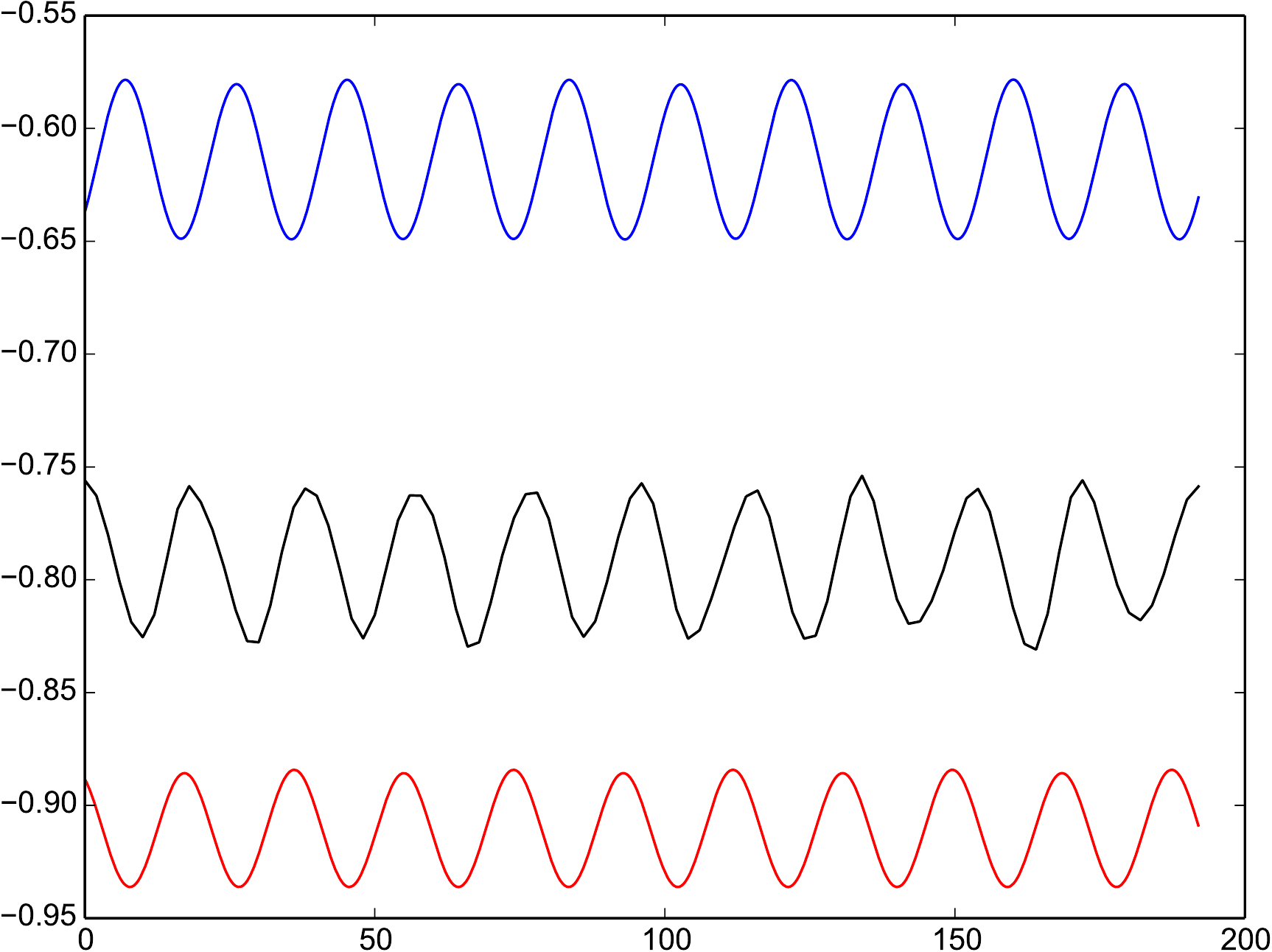}
\end{center}

\caption{Final 10 periods of evolution (approximately 192 time units)
  in the CLAM, GGM and point-vortex simulations.  Time is shown relative
  to $t_{\mathsf{max}}-192$ in each simulation.  Colours (online) show
  CLAM in black, GGM in blue, and the point vortex model in red.}
\label{fig-Wfinal10}
\end{figure}

With $\Gamma_k$ and $\hr_k$ determined in this way, the governing 
dynamical system
\begin{equation}
\dd{{\hr}_k}{t} = \frac{1}{4\pi} \sum_{j=1,\, j\neq k}^4 \Gamma_j 
\frac{\hr_j \times \hr_k}{1-\hr_j \bcdot \hr_k}
\quad (k=1,\,2,\,3,\,4)
\label{pvmodel}
\end{equation}
\citep{polvani93} was integrated forwards for $4000$ time units, 
with the results diagnosed as above for the configuration $W(t)$ 
to compare with the CLAM and GGM results.  The key finding is that
$W(t)$ exhibits the same behaviour, in particular with the same 
dominant frequency of oscillation; the last 10 periods of oscillation
in all three simulations is shown in figure~\ref{fig-Wfinal10}.
(The averages and amplitudes of $W(t)$ differ slightly due to the largely unpredictable nature of turbulence and the crude fitting method of the point-vortex system.)
Much longer point-vortex simulations carried out to $t=10^6$ exhibit 
the same behaviour, with undiminished oscillations.  

\begin{figure}
\begin{center}
\includegraphics[width = 5.24in]{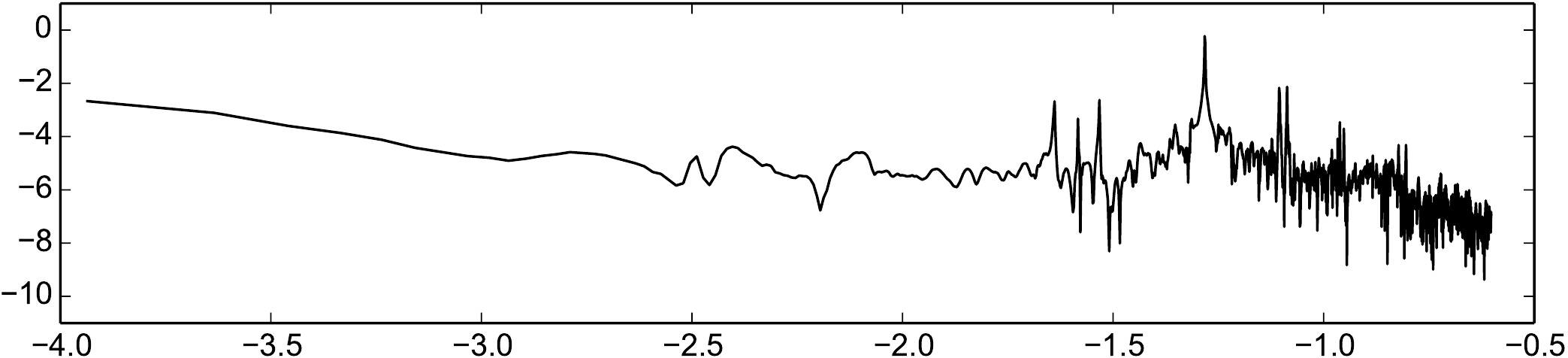} \\
\includegraphics[width = 5.24in]{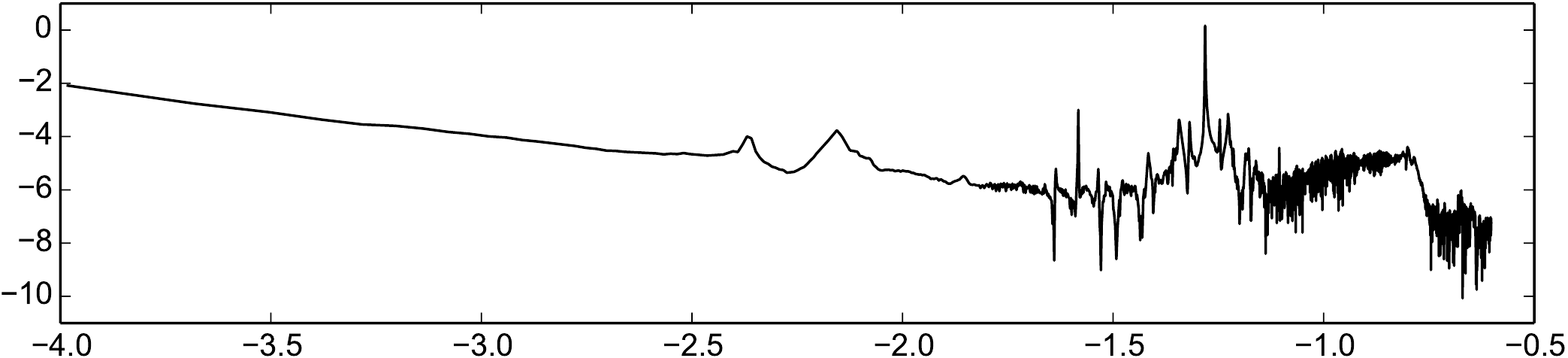} \\
\includegraphics[width = 5.24in]{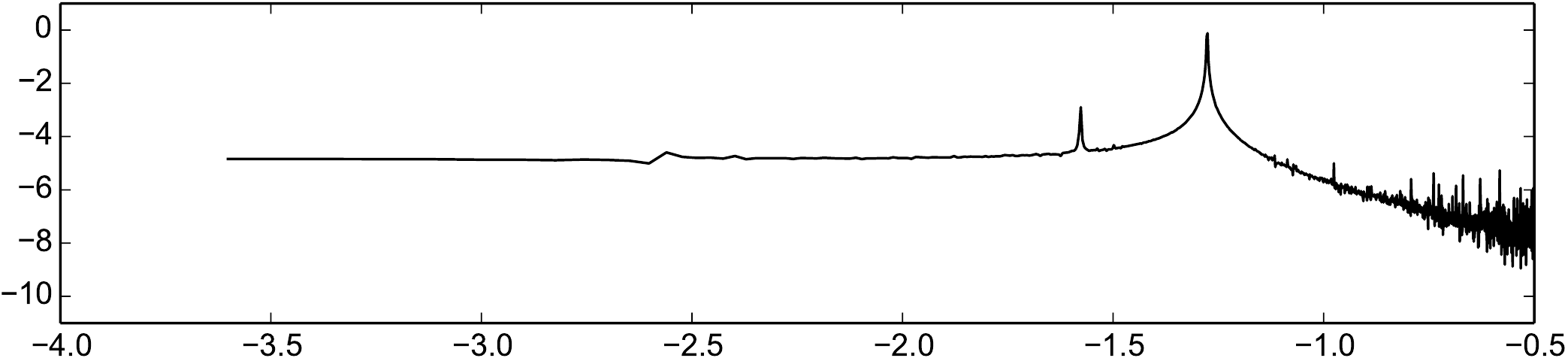}
\end{center}

\caption{Configuration frequency spectrum $\cW(f)$ for 
  the second highest resolution GGM simulation (top), the highest 
  resolution CLAM simulation (middle), and the point vortices (bottom).
  In GGM, $\cW(f)$ was computed from the time series of $W$ for 
  $t \in [400,10000]$; in CLAM we used $[400,9022]$,
  while in the point-vortex model we used $[0,4000]$.
  Note: $\log_{10}\cW$ is plotted versus $\log_{10}f$.}
\label{fig-fspec}
\end{figure}

The correspondence between the three simulations is even more
remarkable if one examines the frequency power spectrum of $W$, namely
$\cW(f)=|\hat{W}|^2$, where $\hat{W}$ is the Fourier transform of $W$
and $f$ is the frequency (here equal to the inverse period).  The
frequency spectra ($\log_{10}$ scaled) for the three simulations are
shown in figure~\ref{fig-fspec}, vertically aligned, with GGM on top, CLAM
in the middle, and the point vortices on the bottom.  
Most strikingly, the dominant frequency $f = 1/T$ with $T \approx 19.14$
is matched in all three simulations.  The power at this frequency is 
orders of magnitude above surrounding frequencies, indicating that the
dynamics are very nearly periodic.  Other less prominent peaks also 
line up, corresponding to longer period oscillations.  Overall, the
agreement is remarkable, given the enormous simplifications made to
reduce the dynamics to that of 4 point vortices.  Most importantly,
the point vortex model strongly supports our claim that the dynamics 
never comes to equilibrium.

\section{Conclusions}

In this paper, we have studied the long-time asymptotic behaviour of a
bounded two-dimensional inviscid flow.  Statistical-mechanical
theories stemming from \citet{miller90,miller92} and
\citet{robert91a,robert91b} argue that this behaviour reaches a steady
equilibrium state, characterised by a functional relation between
vorticity and streamfunction.  We have tested these theories by
carrying out very long time simulations of 2D turbulence on a sphere,
for zero angular momentum, employing highly distinct numerical
methods.  The key finding, which is robust across both methods at all
resolutions, is that the flow remains persistently unsteady with no
indication of equilibration in a steady flow.  

Of course, we cannot carry out simulations indefinitely, and one can
always argue that equilibration will happen eventually.  However, we
have shown there is a plausible alternative.  At long times, the
vorticity field is primarily concentrated in 4 large vortices.  They
are generally unequal in circulation, and not regularly spaced.  If
one replaces each vortex by a point vortex of the same circulation, we
have shown that the point vortices typically exhibit unsteady
quasi-periodic motion --- persistent unsteadiness.  Fundamentally, we
argue, the observed unsteadiness in the Euler equations for continuous
vorticity is intimately associated with the unsteadiness of 4 point
vortices moving on the surface of a sphere. 

We have also found that small-scale vortices pepper the flow field at
high resolution.   These vortices commonly have vorticity anomalies
which cause the vortices to rotate in the same sense as the shear 
induced by differential rotation of the main vortex they circulate 
about.  The small-scale vortices therefore experience
a stabilising cooperative shear \citep{dritschel90}.   As a
result, even these small-scale vortices may persist indefinitely.  Moreover, 
as the resolution increases, more structure at small scales appear.  Our
main conclusion is that bounded inviscid flows on a sphere, with zero
angular momentum, typically do not settle down to an
equilibrium; instead, a few robust compact vortices emerge from the
early flow evolution and interact chaotically or quasi-periodically. 

In the much studied case of doubly-periodic geometry, by contrast,
often two opposite-signed vortices emerge.  In this case, the
corresponding point-vortex system is integrable, and the only possible
motion is simple steady translation.  This difference from spherical
geometry greatly facilitates the emergence of a steady flow (in a
translating frame of reference), though one still cannot rule out the
persistence of small-scale vortices circulating around the main
large-scale vortices.  Inviscid flows in a doubly-periodic domain are likely to
be less unsteady than flows on a sphere, though unsteadiness may
persist forever.

\acknowledgments
We are grateful for helpful discussions with Freddy Bouchet, James Cho, Bruce Turkington, Antoine Venaille, and Peter Weichman.  
We thank the Kavli Institute for Theoretical Physics for supporting our
participation in the 2014 Program ``Wave-Flow Interaction in
Geophysics, Climate, Astrophysics, and Plasmas'' where this work was
initiated.  The KITP is supported in part by the NSF Grant No.\ NSF
PHY11-25915. This work was also supported in part by the NSF under
grant Nos.\ DMR-1306806 and CCF-1048701 (JBM and WQ).

\appendix
\label{append}
\section{Initial Condition}
The initial streamfunction at time $t = 0$ is a superposition of spherical harmonics with spherical wavenumber in the range $4 \leq \ell \leq 10$,
\begin{eqnarray}
\psi(\theta, \phi) = \sum_{\ell = 4}^{10} \sum_{m = -\ell}^\ell \psi_{\ell m}~ Y_\ell^m(\theta, \phi)~, 
\end{eqnarray}
with $\psi_{\ell -m} = (-1)^m \psi_{\ell m}^*$ and the following choice of amplitudes $\psi_{\ell m}$:
\begin{eqnarray}
\psi_{4 0} &=& -0.0114727\nonumber \\
\psi_{4 1} &=& 0.0255012 + 0.0337679~ i\nonumber \\
\psi_{4 2} &=& -0.0366116 - 0.0312419~ i\nonumber \\
\psi_{4 3} &=& 0.000614585 + 0.000602384~ i\nonumber \\
\psi_{4 4} &=& 0.0160784 - 0.0194655~ i\nonumber
\end{eqnarray}
\begin{eqnarray}
\psi_{5 0} &=& 0.00357198\nonumber \\
\psi_{5 1} &=& -0.000446705 - 0.0100286~ i\nonumber \\
\psi_{5 2} &=& 0.0100984 + 0.00778139~ i\nonumber \\
\psi_{5 3} &=& -0.0142904 - 0.00864324~ i\nonumber \\
\psi_{5 4} &=& 0.0156104 - 0.0111872~ i\nonumber \\
\psi_{5 5} &=& -0.00275238 - 0.0106832~ i\nonumber
\end{eqnarray}
\begin{eqnarray}
\psi_{6 0} &=& -0.00352493\nonumber \\
\psi_{6 1} &=& -0.00739953 + 0.00218422~ i\nonumber \\
\psi_{6 2} &=& 0.0140976 + 0.00864871~ i\nonumber \\
\psi_{6 3} &=& -0.0141289 - 0.00654953~ i\nonumber \\
\psi_{6 4} &=& 0.0200989 - 0.0152088~ i\nonumber \\
\psi_{6 5} &=& 0.00682398 + 0.0132045~ i\nonumber \\
\psi_{6 6} &=& 0.00919374 + 0.00889445~ i\nonumber
\end{eqnarray}
\begin{eqnarray}
\psi_{7 0} &=& 0.00269739\nonumber \\
\psi_{7 1} &=& -0.00607359 - 0.00179029~ i\nonumber \\
\psi_{7 2} &=& 0.00299202 - 0.0115332~ i\nonumber \\
\psi_{7 3} &=& -0.00418447 - 0.000665072~ i\nonumber \\
\psi_{7 4} &=& 0.00853673 - 0.0011805~ i\nonumber \\
\psi_{7 5} &=& -0.00229449 - 0.00731682~ i\nonumber \\
\psi_{7 6} &=& 0.00315312 - 0.00332254~ i\nonumber \\
\psi_{7 7} &=& 0.0127047 + 0.000153664~ i\nonumber
\end{eqnarray}
\begin{eqnarray}
\psi_{8 0} &=& -0.0070908\nonumber \\
\psi_{8 1} &=& -0.00820113 - 0.00180399~ i\nonumber \\
\psi_{8 2} &=& 0.00726466 - 0.00109238~ i\nonumber \\
\psi_{8 3} &=& 0.00327615 + 0.00732665~ i\nonumber \\
\psi_{8 4} &=& -0.00418371 + 0.00764604~ i\nonumber \\
\psi_{8 5} &=& -0.00506492 + 0.00152613~ i\nonumber \\
\psi_{8 6} &=& 0.00289222 - 0.0126644~ i\nonumber \\
\psi_{8 7} &=& -0.00113653 - 0.00983086~ i\nonumber \\
\psi_{8 8} &=& 0.00390903 + 0.00337946~ i\nonumber
\end{eqnarray}
\begin{eqnarray}
\psi_{9 0} &=& 0.00366743\nonumber \\
\psi_{9 1} &=& -0.00670701 - 0.0052387~ i\nonumber \\
\psi_{9 2} &=& -0.00102863 - 0.00831521~ i\nonumber \\
\psi_{9 3} &=& -0.00429645 + 0.00881862~ i\nonumber \\
\psi_{9 4} &=& 0.00487287 - 0.00708109~ i\nonumber \\
\psi_{9 5} &=& -0.00183309 - 0.00476393~ i\nonumber \\
\psi_{9 6} &=& 0.0124893 - 0.00170479~ i\nonumber \\
\psi_{9 7} &=& 0.00438044 + 0.00091529~ i\nonumber \\
\psi_{9 8} &=& -0.00309726 - 0.000427295~ i\nonumber \\
\psi_{9 9} &=& -0.0053439 + 0.00203709~ i\nonumber
\end{eqnarray}
\begin{eqnarray}
\psi_{10 0} &=& -0.00507863\nonumber \\
\psi_{10 1} &=& -0.00247735 + 0.00024564~ i\nonumber \\
\psi_{10 2} &=& -0.00357194 - 0.00141963~ i\nonumber \\
\psi_{10 3} &=& -0.00077508 + 0.00719427~ i\nonumber \\
\psi_{10 4} &=& -0.00462811 - 0.00550109~ i\nonumber \\
\psi_{10 5} &=& -0.00101044 - 0.00773111~ i\nonumber \\
\psi_{10 6} &=& 0.00395564 + 0.0027653~ i\nonumber \\
\psi_{10 7} &=& 0.00121852 + 0.00310834~ i\nonumber \\
\psi_{10 8} &=& 0.00314655 + 0.000289766~ i\nonumber \\
\psi_{10 9} &=& -0.00279021 - 0.00314691~ i\nonumber \\
\psi_{10 10} &=& 0.00258861 - 0.000254598~ i
\end{eqnarray}

\bibliographystyle{jfm}

\bibliography{late2d}

\end{document}